\documentclass[prd,nofootinbib,12pt]{revtex4}

\usepackage{lineno,hyperref}
\usepackage{amsfonts,amssymb,amsmath}
\usepackage{epsfig}
\usepackage{mathrsfs}
\usepackage{amssymb}
\usepackage{graphicx}
\usepackage{amsmath}
\usepackage{graphicx}
\usepackage{color}

\begin{document}

\title{ Krylov Complexity in early universe}
\author{Ke-Hong Zhai$^{1}$}
\email{2023700328@stu.jsu.edu.cn}
\author{Lei-Hua Liu$^{1}$}
\email{liuleihua8899@hotmail.com}

\affiliation{Department of Physics, College of Physics,
	Mechanical and Electrical Engineering,
	Jishou University, Jishou 416000, China
}

\begin{abstract}
	
The Lanczos algorithm offers a framework for constructing wave functions in closed and open quantum systems from their Hamiltonians. Since the early universe is inherently an open system, we employ this algorithm to investigate Krylov complexity across various cosmological phases: inflation, radiation domination (RD), and matter domination (MD). Our results highlight a clear distinction in Krylov complexity between the closed- and open-system methodologies. To accurately capture the influence of potentials during RD and MD, we examine a set of inflationary potentials, including the Higgs potential, $R^2$ inflation, and chaotic inflation, while incorporating violations of slow-roll conditions. This study is conducted in conformal time through the preheating stage. Numerically, we find that the evolution of Krylov complexity and Krylov entropy shows remarkable similarity across different potentials during RD and MD. Furthermore, we rigorously construct an open two-mode squeezed state using the second kind of Meixner polynomial. Based on this construction, we derive for the first time the evolution equations for the squeezing parameter $r_k$ and phase $\phi_k$ in terms of the scale factor. Our analysis indicates that dissipative effects lead to rapid decoherence-like behavior. In addition, we observe that the inflationary universe behaves as a strongly dissipative system, whereas during the RD and MD epochs the universe exhibits weak dissipative characteristics. This work opens new perspectives for studying the universe from a quantum-informational viewpoint.

\end{abstract}

\maketitle

\tableofcontents

\section{Introduction}
\label{sec:intro}

The concept of complexity has become increasingly important in high-energy physics, although a unified definition is still lacking. Nonetheless, several computational frameworks have been proposed to evaluate it. One major approach is computational complexity, which quantifies the minimum number of logical steps required to perform a given task \cite{Aaronson:2016vto}. Another significant line of research stems from Nielsen and collaborators, who introduced a geometric method for assessing complexity \cite{Nielsen:2005mkt,Dowling:2006tnk,Nielsen:2006cea}. A related framework connects complexity to the Fubini–Study distance in the geometry of information space \cite{Chapman:2017rqy}. However, both geometric approaches depend strongly on the choice of metric on the underlying parameter manifold.

A more recent development is the notion of Krylov complexity, which circumvents the need to construct a parameter manifold altogether \cite{viswanath1994recursio}. By operating directly in the operator Hilbert space, this approach avoids certain ambiguities inherent in geometric formulations, though the precise connections and distinctions between Krylov complexity and other frameworks remain an active area of research. Within this paradigm, the concept of universal operator growth has been proposed \cite{Parker:2018yvk}, suggesting that the late-time dynamics of quantum operators exhibit universal features encoded in their Krylov complexity. Given a specific quantum operator, one can explicitly compute its Krylov complexity, associated Krylov entropy, and the sequence of Lanczos coefficients using the recursive Lanczos algorithm \cite{viswanath1994recursio}. The relationship between Krylov complexity and circuit complexity has been examined from contrasting viewpoints: while Ref. \cite{Aguilar-Gutierrez:2023nyk} argues that the two notions are generally incompatible. However, Ref. \cite{Caputa:2021sib} shows that, under certain conditions, Krylov complexity is proportional to the distance measured by the Fubini–Study metric.

Recently, Krylov complexity has found broad applications across both condensed matter and high-energy physics. A systematic study employing various orthogonal polynomial bases was carried out in Ref.~\cite{Muck:2022xfc}, highlighting the flexibility and robustness of the Krylov framework. Given the central role of the Sachdev–Ye–Kitaev (SYK) model in understanding quantum chaos and holography, Krylov complexity has been actively explored in this context as well \cite{Rabinovici:2020ryf,Jian:2020qpp,He:2022ryk}. Beyond the SYK model, it has been applied to the analysis of generalized coherent states \cite{Patramanis:2021lkx}, spin systems such as the Ising and Heisenberg models \cite{Cao:2020zls,Trigueros:2021rwj,Heveling:2022hth}, and conformal field theories \cite{Dymarsky:2021bjq,Caputa:2021ori,Kundu:2023hbk}. More recently, the formalism has been extended to topological phases of matter \cite{Caputa:2022eye} and even to quantum field theory \cite{He:2024hkw,He:2024xjp}, underscoring its growing relevance as a universal diagnostic of operator dynamics and quantum complexity. 

It was shown in Ref.~\cite{Dymarsky:2019elm} that quantum chaos can be interpreted as delocalization in Krylov space, an insight that has reshaped the understanding of operator growth in chaotic systems. Interestingly, exponential growth of Krylov complexity has also been observed in certain integrable models exhibiting saddle-dominated scrambling, with universal features confirmed in Ref.~\cite{Huh:2023jxt}. Since then, the scope of Krylov complexity has continued to expand rapidly. Recent developments include its application to open quantum systems, non-Hermitian dynamics, holography, random matrix theory, and out-of-equilibrium phenomena—see, e.g., Refs.~\cite{Erdmenger:2023wjg,Hashimoto:2023swv,Vasli:2023syq,Gill:2023umm,Bhattacharjee:2023uwx,Adhikari:2022whf,Loc:2024oen,Caputa:2024vrn,Basu:2024tgg,Sasaki:2024puk,Caputa:2024xkp,Bhattacharjee:2022qjw,Sahu:2024opm,Kim:2021okd,Chen:2024imd,Bhattacharjee:2024yxj,Sanchez-Garrido:2024pcy,Chattopadhyay:2024pdj,Balasubramanian:2024ghv,Bhattacharya:2024hto,Mohan:2023btr}. Notably, the framework has also been generalized beyond standard unitary quantum mechanics, with extensions to stochastic dynamics, classical systems, and information-theoretic settings \cite{Nandy:2024htc}.

In this work, we aim to investigate Krylov complexity throughout the entire evolution of the early universe, spanning the inflationary epoch, the subsequent RD era, and the MD era. Previous studies have explored computational complexity during inflation \cite{Choudhury:2020hil,Bhargava:2020fhl,Lehners:2020pem,Bhattacharyya:2020rpy,Adhikari:2021ked}, uncovering an oscillatory behavior at the onset of inflation and a qualitatively similar pattern in the post-inflationary phase. When quantum effects are incorporated through various quantum-information-theoretic frameworks, the complexity displays irregular oscillations during inflation, a feature that provides a novel diagnostic for distinguishing between competing inflationary models \cite{Li:2023ekd}.

While several recent works have begun to explore Krylov complexity in the context of the early universe, Ref.~\cite{Adhikari:2022oxr} stands out as the first to investigate it during inflation, specifically accounting for non-trivial sound speed effects within a closed quantum system framework. However, given that the early universe is more realistically modeled as an open quantum system, interacting with unobservable environmental degrees of freedom, a more physically complete treatment requires going beyond unitary evolution. Addressing this, Ref.~\cite{Li:2024kfm} adopted an open quantum system approach to study Krylov complexity under a modified dispersion relation during inflation. Their analysis reveals that inflation behaves as a strongly dissipative process, with Krylov complexity displaying a persistent overall increase. A crucial ingredient in this open-system formulation is the explicit construction of the time-evolving wave function, as systematically developed in Ref.~\cite{Bhattacharya:2022gbz}. In the present work, we take the wave function derived in Ref.~\cite{Li:2024kfm} as our starting point to extend the analysis across the full cosmological timeline, from inflation through RD and MD eras.

In previous studies, Krylov complexity has been applied to the early universe using a thermal state as the reference. Refs.~\cite{Li:2024ljz, Li:2024iji} extended this framework by incorporating the contribution of the scalar potential—approximated by its quadratic term, and examined the impact of slow-roll violation on complexity growth. In our earlier work~\cite{Li:2024ljz}, the potential was expanded quadratically around its minimum; while analytically convenient, this approximation may not fully capture the dynamics of realistic inflationary scenarios.

It has been suggested by Ref.~\cite{Galante:2014ifa} that many inflationary models are unified into the framework of $\alpha$-attractors, whose potential form closely resembles that of $R^2$ inflation. Moreover, Ref.~\cite{Albrecht:1992kf} established a connection between the Bunch-Davies vacuum and the two-mode squeezed state. Motivated by these two key insights, we will continue to employ the two-mode squeezed state formalism to study Krylov complexity in the context of the entire early universe. To improve physical fidelity, we plan in upcoming work to generalize the analysis to well-motivated inflationary potentials, including the Higgs potential~\cite{Higgs:1964pj}, Starobinsky’s inflation~\cite{Starobinsky:1980te}, and chaotic inflation~\cite{Linde:1983gd}. These models encompass most inflationary models and are especially relevant in light of current observational constraints, which strongly favor plateau-like potentials. This extension will allow for a more accurate and observationally grounded exploration of quantum complexity during inflation.

This paper is organized as follows. In Sec.~\ref{Lanczos algorithm}, we will introduce the Lanczos algorithm. Sec.~\ref{section some basics of early universe} will provide a brief review of the early universe, covering inflation, RD, and MD eras, along with three representative inflationary potentials. In Sec.~\ref{section The evolution of}, we will present the two-mode squeezed state formalism and numerically solve for the parameter $r_k$ across different cosmological periods. Sec.~\ref{section Krylov complecity} will investigate the Krylov complexity in the framework of a closed-system methodology, which Sec.~\ref{section Krylov entropy} will examine the corresponding Krylov entropy. In Sec.~\ref{section op krylov complexity}, we will analyze the Lanczos coefficients and Krylov complexity using an open-system methodology; notably, we will present  the evolution equations for $\phi_k$ and $r_k$ derived within the open two-mode squeezed state formalism in the first time. Finally, conclusions and outlook will be given in Sec.~\ref{section summary and outlook}.

In this work, we will work in Planck units, namely $c=M_{P}=G=1$.


\section{Lanczos algorithm with  closed-system  methodology}
\label{Lanczos algorithm}
In this section, we will mainly review the Lanczos algorithm with closed-system  methodology. First, the operator $\mathcal{O}(t)$ in Heisenberg picture needs to be defined as follows,
\begin{equation}
\partial_t \mathcal{O}(t)=i[H,\mathcal{O}(t)]
\label{operator O}
\end{equation}
where H is the Hamiltonian and its solution is
\begin{equation}
\mathcal{O}(t)=e^{iHt}\mathcal{O}e^{-iHt}
\label{solution of O}
\end{equation}
Next, let's define the Liouvillian super-operator $\mathcal{L}_X$ as $\mathcal{L}_X Y=[X,Y]$, where $[,]$ denotes the commutator. We apply this to Eq. \eqref{solution of O}, the operator is also expressed as 
\begin{equation}
\mathcal{O}(t)=e^{i\mathcal{L}t}\mathcal{O}=\sum_{n=0}^{\infty} \frac{(it)^n}{n!}\mathcal{L}^{n}\mathcal{O}(0)=\sum_{n=0}^{\infty} \frac{(it)^n}{n!}\Tilde{\mathcal{O}}_{n}
\label{O with L}
\end{equation}
where $\mathcal{L}^{n}\mathcal{O}=\Tilde{\mathcal{O}}_{n}=[H,\Tilde{O}_{n-1}]$ which can serve as the basis for the Hilbert space.
\begin{equation}
\begin{split}
\mathcal{O}\equiv |{\tilde{\mathcal{O}} })  ,\mathcal{L}^{1}\mathcal{O}\equiv |{\tilde{\mathcal{O}} }_{1})  ,\mathcal{L}^{2}\mathcal{O}\equiv |{\tilde{\mathcal{O}} }_{2})  ,\mathcal{L}^{3}\mathcal{O}\equiv |{\tilde{\mathcal{O} } }_{3}) ...      
\end{split}
\label{basis of O}
\end{equation}
However, they are not orthogonal bases. In order to construct an orthogonal basis, we use the Lanczos algorithm. The first two operators become
\begin{equation}
\mathcal{O}_{0}=|\Tilde{\mathcal{O}}_{0})=\mathcal{O}, ~~~|\mathcal{O}_{1})=b_{1}^{-1}\mathcal{L}|\mathcal{O}_{0})
\label{first two O}
\end{equation}
where $b_{1}=\sqrt{(\Tilde{\mathcal{O}}_{0}\mathcal{L}|\mathcal{L}\Tilde{\mathcal{O}}_{0})}$ describing the normalized vector. And the next orthogonal basis is written by
\begin{equation}
|\mathcal{O}_{n})=b_{n}^{-1}|A_{n})
\label{n-th turns O}
\end{equation}
with
\begin{equation}
|A_{n})=\mathcal{L}|\mathcal{O}_{n-1})-b_{n-1}|\mathcal{O}_{n-2}),~~~b_{n}=\sqrt{(A_{n}|A_{n})}
\label{n-th A}
\end{equation}
where $b_{n}$ represents the Lanczos coefficients. This iterative relation will stop and produce the finite orthogonal Krylov basis if $b_{n}=0$. Subsequently, we express Eq. \eqref{O with L} as follows,
\begin{equation}
\mathcal{O}(t)=e^{i\mathcal{L}t}\mathcal{O}=\sum_{n=0}^{\infty} (i)^{n}\phi_{n}(t)|\mathcal{O}_{n})
\label{O with O_n}
\end{equation}
where $\phi_{n}$ is the wave function satisfying with $\sum_{n} |\phi_{n}|^{2}=1$. By substituting Eq. \eqref{O with O_n} into the Schr$\Ddot{o}$dinger equation, we obtain
\begin{equation}
\partial_{t}\phi_{n}(t)=b_{n}\phi_{n-1}-b_{n+1}\phi_{n+1}
\label{O in SE}
\end{equation}
Next, we define the Krylov complexity using the wave function $\phi_{n}$ as follows
\begin{equation}
K=\sum_{n} n|\phi_{n}|^{2}
\label{definition of KC}
\end{equation}
Additionally, the Lanczos coefficient is bounded via \cite{Parker:2018yvk}, 
\begin{equation}
b_{n}\le \alpha n+\eta
\label{boundary of b_n}, 
\end{equation}
where $\alpha$ and $\eta$ encode different information in various models. Note that $b_n = \alpha n + \eta$, which indicates the system is a maximally chaotic dynamical system. For the case of maximal chaos, we obtain the relationship between the Lyapunov exponent and $\gamma$ and $\alpha$ as follows
\begin{equation}
\lambda=2\alpha
\label{chaotic growth}
\end{equation}
In this work, we will investigate the Krylov complexity for the entire early universe, thus the formula of scale factor $a(\eta)$ should be introduced.

\section{Some setup of early universe}
\label{section some basics of early universe}
In this section, we first review some fundamental aspects of the scale factor in different cosmological epochs. Following the notation of our previous work \cite{Li:2024ljz}, a central goal of this study is to examine the impact of the inflationary potential on Krylov complexity during the RD and MD eras, particularly due to the violation of slow-roll conditions. The Mukhanov-Sasaki variable incorporates contributions from the potential in a way that does not explicitly reveal its influence on Krylov complexity. Therefore, in this work, we perturb the inflaton field directly. We then introduce several important inflationary potentials: the chaotic inflationary potential, the Starobinsky potential ($R^2$ inflation), and the Higgs potential. Finally, we employ preheating dynamics to numerically evaluate the effective mass of the inflaton with respect to the scale factor.

\subsection{Inflation, RD and MD}
We will follow the notation of \cite{Baumann:2009ds} to introduce the scale factor. The background metric is the Friedman-Lemaitre-Robertson-Walker (FLRW) metric,
\begin{equation}
ds^2=a^2(\eta)(-d\eta^2+d\vec{x}^2),
\label{bacground metric}
\end{equation}
where $a(\eta)$ is the scale factor expressed in conformal time $\eta$. Alternatively, the FLRW metric can also be written in terms of physical time $t$ as $ds^2 = -dt^2 + a(t)^2 d\vec{x}^2$, from which the relation $dt = a d\eta$ follows explicitly. Here, we consider a universe dominated by a single component during each epoch, meaning it undergoes successive phases of inflation, RD, and MD. The evolution of the scale factor in each era can be characterized by the equation of state parameter $w$ via 
\begin{equation}
w_I=\frac{P_I}{\rho_I},
\label{equation of state}
\end{equation}
where $P_I$ and $\rho_I$ are the pressure and energy density, respectively, in each epoch $I$. To derive the relation between conformal time and the scale factor, we make use of the comoving distance,
\begin{equation}
\chi_{ph}(\eta)=\int_{t_{i}}^{t} \frac{dt}{a}=\int_{\ln a_{i}}^{\ln a}(aH)^{-1}d\ln a
\label{eq. comoving distance}
\end{equation}
where $\chi_{\mathrm{ph}}(\eta)$ is the comoving particle horizon distance, and we have used the definition of the Hubble parameter $H = \dot{a}/a$, with $\dot{a} = da/dt$. In terms of the equation of state parameter $w_I$, the Hubble radius can be expressed as
\begin{equation}
(aH)^{-1}=H_{0}^{-1}a^{\frac{1}{2}(1+3\omega)}
\end{equation}
where, for simplicity, we replace $\omega_{I}$ with $\omega$. Integrating Eq. \eqref{eq. comoving distance} yields an explicit functional relation between the scale factor and the conformal time, 
\begin{equation}
\eta=\frac{2H_{0}^{-1}}{(1+3\omega)}a^{\frac{1}{2}(1+3\omega)}. 
\label{relation eta and a}
\end{equation}
Consequently, we derive the relation for various time periods
\begin{equation}
\eta=\begin{cases}
& -(aH_{0})^{-1},  \   \ (\omega=-1), \ \  \rm inflation \\
& aH_{0}^{-1}, \  \ (\omega=\frac{1}{3}), \ \ \rm RD \\
& 2\sqrt{a}H_{0}^{-1}, \  \ (\omega=0), \ \ \rm MD
\end{cases}
\label{formula of eta}
\end{equation}
It shows that the various values of $\omega$ describe the distinctive periods.

In the following, we outline the fundamental evolution of the universe across its distinct epochs. The universe emerges from a singularity and undergoes a phase of exponential expansion known as inflation. During this period, curvature perturbations transition from quantum to classical after horizon exit, becoming imprinted on the Cosmic Microwave Background (CMB). Subsequently, the universe enters the RD era, characterized by an energy density $\rho_{\rm rad}(t) = \rho_{\rm rad_0} , a_{0}^{4} / a(t)^{4}$, where $\rho_{\rm rad_0}$ and $a_{0}$ denote the initial energy density and scale factor at the onset of RD. In this epoch, particles are relativistic. As the temperature drops, particles become increasingly non-relativistic, leading the universe into the MD era, where the energy density evolves as $\rho_{\rm mat}(t) = \rho_{\rm mat_0} , a_{0}^{3} / a(t)^{3}$. The transition from RD to MD occurs when the energy densities of radiation and matter equalize. During the subsequent MD epoch, a preheating process can generate particles. We begin by presenting the full action for single-field inflation:
\begin{equation}
S=\int d\eta d^3\vec{x}\sqrt{-g}\bigg[\frac{1}{2}g^{\mu\nu}\partial_\mu\phi\partial_\nu\phi-V(\phi)\bigg],
\label{total action}
\end{equation}
where $V(\phi)$ is the potential of inflation. Following the method of \cite{Kofman:1997yn,Shtanov:1994ce}, the equation of motion for background $\phi(t)$ in RD and MD is 
\begin{equation}
\Ddot{\phi}(t)+3H\dot{\phi}(t)+V(\phi),_{\phi}=0,
\label{eom of phit}
\end{equation}
where $\Ddot{\phi}(t)=\partial_t^2 \phi(t)$, $V(\phi),_{\phi}=\frac{dV(\phi)}{d\phi}$ and the similar work is found in our previous works \cite{Liu:2019xhn,Liu:2021rgq}. 
Defining with the effective mass as follows,
\begin{equation}
m_{\rm eff}=\frac{d^2V}{d\phi^2}=V_{,\phi\phi}. 
\label{effective mass}
\end{equation}
Here, we explain why the effective mass is defined as $m_{\rm eff} = d^2V/d\phi^2$. For the different inflationary potentials in Eq. \eqref{eq Vi}, we take $V_{\rm chaotic}$ as an example, where one can directly show that $m_{\rm eff}^2 = m^2$. Based on this, we adopt the same definition of effective mass for all inflationary potentials considered. To solve Eq. \eqref{eom of phit}, we introduce a new variable $\tilde{\phi}(t) = a^{3/2} \phi(t)$, which transforms Eq. \eqref{eom of phit} into
\begin{equation}
\Ddot{\tilde{\phi}}(t)+m_{\rm eff}^2\tilde{\phi}(t)=0,
\label{eom of phit1}
\end{equation}
The solution to Eq. \eqref{eom of phit1} is $\phi(t) \propto a^{-3/2} \cos(m_{\rm eff} t)$, which describes the behavior of $\phi(t)$ during the preheating period. Here, the terms $\frac{9}{4} H^2 \tilde{\phi} + \frac{3}{2} \dot{H} \tilde{\phi}$ have been neglected, as they are significantly smaller than the potential contribution. Note that this solution is expressed in physical time $t$. For the subsequent analysis in this paper, we will transform it into conformal time using $d\eta = dt/a$, as given by Eq. \eqref{formula of eta}.

\subsection{Various inflationary potentials}
In this subsection, we will use Eq. \eqref{eom of phit1} to establish a relation between the scale factor and various inflationary potentials. There exist many inflationary models, each characterized by a distinct potential. Current observational data favor concave potentials \cite{Planck:2018jri}, a category that includes the potential of $R^2$ inflation. Furthermore, $R^2$ inflation remains consistent with observations from inflationary to large scales. From a theoretical perspective, many inflationary models can be unified within the framework of $\alpha$-attractor models \cite{Galante:2014ifa}, which also encompass $R^2$ inflation. Motivated by these two considerations, we include $R^2$ inflation in our analysis. According to the Standard Model, the Higgs field is the only fundamental scalar field in nature; therefore, the Higgs potential serves as another natural choice \cite{Higgs:1964pj}. Lastly, we consider chaotic inflation with a quadratic potential \cite{Linde:1983gd}. Collectively, these choices cover a representative range of inflationary models.

We consider the following inflationary potentials: 
\begin{equation}
\begin{cases}
&\ V_{\rm chaotic}=\frac{1}{2}m^{2}\phi^{2}, \\
&\ V_{\rm R^2}=\frac{3}{4}M^{2}M_{p}^{2}(1-e^{-\sqrt{\frac{2}{3}}\frac{\phi}{M_{p}}})^{2}, \\
&\ V_{\rm Higss}=\frac{1}{2}m^{2}\phi^{2}+\frac{1}{4}\lambda\phi^{4},
\end{cases}
\label{eq Vi}
\end{equation}
where $M_{p}=(8\pi G)^{-\frac{1}{2}}$ is the reduced Planck mass. Here, $M$ denotes an energy scale whose value is fixed by the observed amplitude of the primordial power spectrum. Using the solution of Eq. \eqref{eom of phit} and the relation $dt = a d\eta$, we can express $V,_{\phi\phi}$ as
\begin{equation}
\begin{cases}
&\ V_{\rm chaotic,\phi\phi}=m^{2},\\
&\ V_{\rm R^2,\phi\phi}=M^{2}\Bigl (-e^{-\sqrt{\frac{2}{3}}\frac{Aa^{-\frac{3}{2}}\sin(m_{\rm eff}a\eta)}{M_{p}}}+2e^{-2\sqrt{\frac{2}{3}}\frac{Aa^{-\frac{3}{2}}\sin(m_{\rm eff}a\eta)}{M_{p}}}\Bigr ),\\
&\ V_{\rm Higss,\phi\phi}=m^{2}+3\lambda A^{2}a^{-3}\sin^{2}(m_{\rm eff}a\eta)
\end{cases}
\label{eq vi phi phi}
\end{equation}
where $\phi(t) = A a^{-\frac{3}{2}} \sin(m_{\text{eff}} t)$ and we have used $dt = a d\eta$. The definition of effective mass in Eq. \eqref{effective mass} has been employed, and clearly $m_{\text{eff}} = \sqrt{V_{i,\phi\phi}}$ in our work ($\rm i=chaotic, R^2, Higgs$). To simplify the subsequent numerical simulations, we define the variable $y=\log_{10}a$. To numerically obtain the effective mass of the distinct inflationary models $V_{\rm R^2}$ and $V_{\rm Higgs}$, iterative evaluations in terms of $\log_{10}a$ are required, based on Eq. \eqref{eq vi phi phi}. Let us clarify the concept of iterations for the effective mass. The Newton-Raphson method is adopted for its computational efficiency and fast convergence rate. In \eqref{eq vi phi phi}, the effective mass \( m_{\rm eff} \) defined by \eqref{effective mass} carries the same physical meaning as the left-hand side of \eqref{eq vi phi phi}. Thus, the relation \eqref{eq vi phi phi} naturally lends itself to an iterative scheme, requiring an initial value \( m_0 \) for the effective mass. This value (\( m_0^2 = 10^{-12} \)) can be fixed by the COBE normalization \cite{Planck:2015sxf}. In our implementation, we perform 50 iterations for the effective mass, meaning that for each single value of \( y \), the calculation is repeated 50 times.
	
	To check the convergence of $V_{\rm R^2,\phi\phi}$ and $V_{\rm Higgs,\phi\phi}$ within the defined interval, we performed iterations at multiple sampling points from $\log_{10}a = 0$ to $\log_{10}a = 6$ with a step size of $0.05$ (this range covers from RD to MD). We find that all points in the range $\log_{10}a = 0$ to $\log_{10}a = 6$ are convergent.
\begin{figure}
	\centering
	\includegraphics[width=0.35\linewidth]{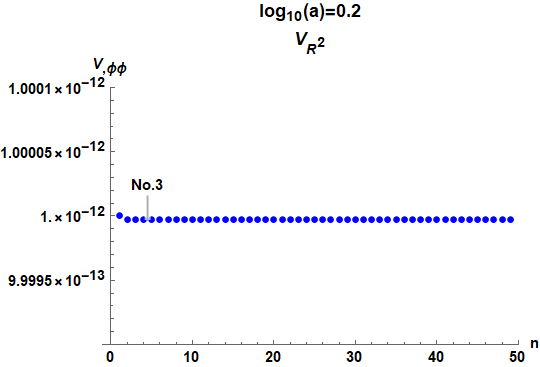}
	\qquad
	\includegraphics[width=0.35\linewidth]{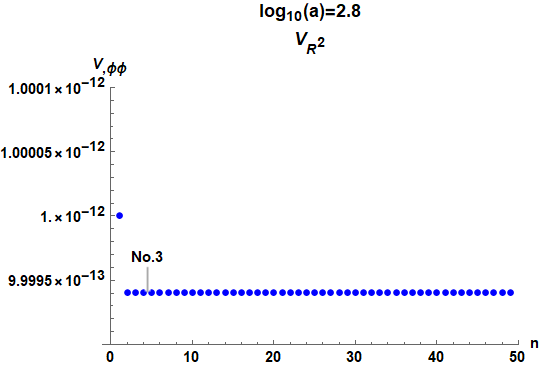}\\
	\qquad
	\includegraphics[width=0.35\linewidth]{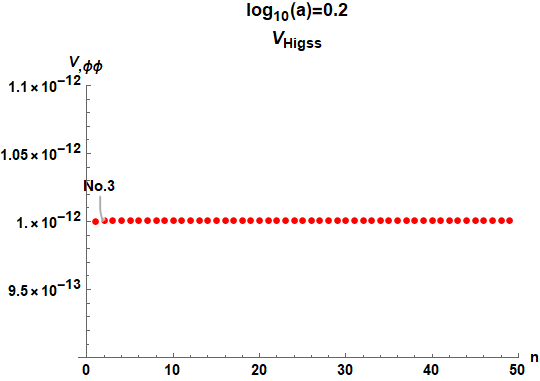}
	\qquad
	\includegraphics[width=0.35\linewidth]{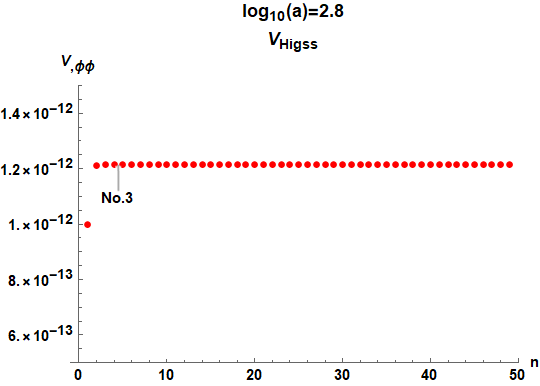}
	\caption{The numeric of $V_{,\phi\phi}$ at first 50 iterations of $V_{\rm R^2,\phi\phi}$ and $V_{\rm Higgs,\phi\phi}$ in $\log_{10}a=0.2$ and $\log_{10}a=2.8$, in which we set $M_{p}=1$, $m=10^{-6}$; $M^{2}=10^{-12}$ at $V_{\rm R^2}$, and $\lambda A^{2}=10^{-4}$ at $V_{\rm Higss}$.}
	\label{fig: v1 v2}
\end{figure}
As an illustration, Fig. \ref{fig: v1 v2} shows the iterative results for  $V_{\rm R^2,\phi\phi}$ and $V_{\rm Higgs,\phi\phi}$, where 50 iterations were performed for two specific values of $\log_{10}a$. The parameters used in the figure are chosen as follows: the potential amplitude is constrained by the COBE normalization \cite{Planck:2015sxf}, which bounds its value below $10^{-10}$ in Planck units, while the inflationary field is a large field with a maximum value around $100$ in these units. Consequently, the effective mass for the various potentials is of order $10^{-6}$, and all other parameters are set consistently with this scaling.

\begin{figure}
	\centering
	\includegraphics[width=0.4\linewidth]{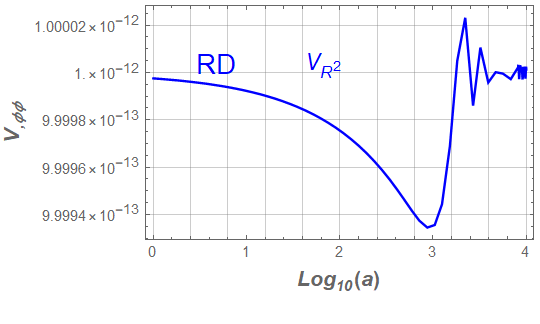}
	\qquad
	\includegraphics[width=0.4\linewidth]{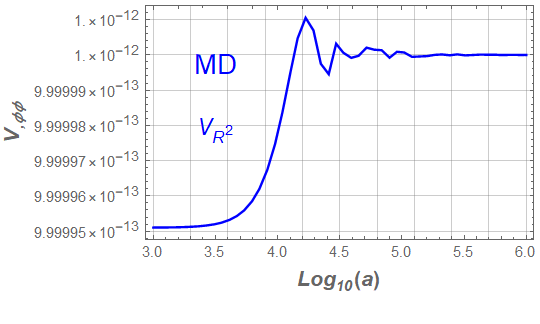}\\
	\qquad
	\includegraphics[width=0.4\linewidth]{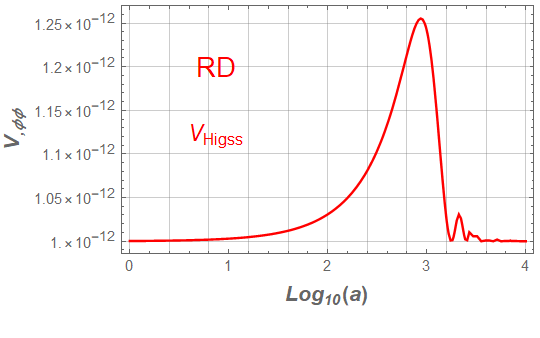}
	\qquad
	\includegraphics[width=0.4\linewidth]{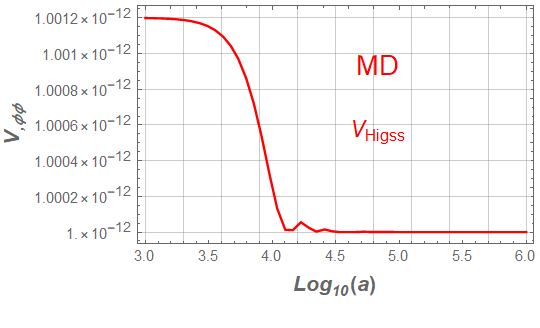}
	\caption{The numeric of $V_{,\phi\phi}$ in terms of $\log_{10}a$ for potential $V_{\rm R^2}$, and $V_{\rm Higgs}$ in RD and MD, in which we replace its exact value  with the iteration of $V_{\rm R^2}$, and $V_{\rm Higgs}$. We set  $M_{p}=1$, $m=10^{-6}$; $M^{2}=10^{-12}$ for $V_{\rm R^2}$, and $\lambda A^{2}=10^{-4}$ for $V_{\rm Higgs}$.}
	\label{fig:RD MD v2 v3}
\end{figure}

Let us now discuss the effects of the distinctive parameters of the various potentials.  
For \( V_{\text{chaotic}} \), its effective mass is explicitly set by the inflaton mass, whose value can be fixed via the COBE normalizations.  
As for \( V_{R^2} \), it contains one parameter \( M \), which acts as the amplitude of the effective mass in Eq. \eqref{eq vi phi phi}; consequently, the range of \( M \) is also constrained by the COBE normalization.  
Finally, the Higgs potential \( V_{\text{Higgs}} \) involves three parameters: the Higgs mass \( m \), the coupling constant \( \lambda \), and the amplitude of the inflaton during preheating \( A \).  
Again, using the COBE normalizations, we can determine the upper bounds on \( m^2 \) and \( \lambda A^2 \); values exceeding these orders would violate the observational constraints.  
Thus, appropriate parameter values can be set from observational data. As illustrated in Fig. \ref{fig:RD MD v2 v3}, the effective masses for all the considered potentials lie within \( 10^{-12} \) in Planck units.

Fig. \ref{fig:RD MD v2 v3} shows the evolution of $V_{,\phi\phi}$ with respect to $\log_{10}a$ during the RD and MD eras, based on the iterative results for $V_{\rm R^2}$ and $V_{\rm Higgs}$ from Fig. \ref{fig: v1 v2}. The evolution of the inflationary phase itself is neglected here due to the violation of slow-roll conditions in these periods. From Fig. \ref{fig:RD MD v2 v3}, we observe that the effective mass for $R^2$ inflation ($V_2$) decreases during RD and then grows, with oscillations, toward a constant value of approximately $10^{-12}$. The behavior for the Higgs potential ($V_{\rm Higgs}$) is different: it exhibits a peak during RD, after which it decreases and settles into an approximately constant value. Notably, the effective masses for both $V_{\rm R^2}$ and $V_{\rm Higgs}$ are of nearly the same order of magnitude, consistent with the COBE normalization constraint \cite{Planck:2015sxf}.

In this section, we introduce the inflationary, RD, and MD eras, each described by its corresponding scale factor. We also present three representative inflationary potentials that cover a broad range of model types. An overview of preheating is given, during which the inflaton background oscillates around its potential minimum. Finally, we incorporate this background solution into a numerical evaluation of the effective mass for each of the three potentials.

\section{Evolution of two-mode squeezed state }
\label{section The evolution of}

As we know, the curvature perturbation will become classical as the horizon exits. Ref. \cite{Grishchuk:1990bj} already shown the amplification from quantum level to classical level during inflation is naturally a quantum squeezing process. Thus, the two-mode squeezed state is quite a natural choice for investigating the early universe. In this section, we will follow the method of \cite{Li:2024kfm} to investigate the evolution of $r_k$ and $\phi_k$ during inflation, RD and MD.

\subsection{Two-mode squeezed state}
\label{title Two-mode squeezed state}
 First, we will simply recall the two-mode squeezed state. To obtain the two-mode squeezed state, we introduce the unitary operator as follows,
\begin{equation}
\mathcal{U}_{k}=\hat{\mathcal{S}}_{k}(\phi_k,r_k)\hat{\mathcal{R}}_{k}(\theta_k)
\end{equation}
with 
\begin{equation}
\hat{\mathcal{S}}_{k}=\exp\bigg[r_k(\eta)\big(e^{-2i \phi_k(\eta)}\hat{c}_{\vec k}\hat{c}_{-\vec k}-e^{2i \phi_k(\eta)}\hat{c}^{\dagger}_{-\vec k}\hat{c}_{\vec k}^\dagger\big)\bigg]  
\end{equation}
\begin{equation}
\hat{\mathcal{R}}_{k}(\theta_k)=\exp\bigg[-i\theta_k(\eta)\big(\hat{c}_{\vec k}\hat{c}_{\vec k}^\dagger+\hat{c}_{-\vec k}^\dagger\hat{c}_{-\vec k}\big)\bigg]
\end{equation}
rwhere $\hat{\mathcal{S}}_{k}$ is the squeeze operator and $\hat{\mathcal{R}}_{k}(\theta_k)$ is the rotation operator, which can be neglected as it only contributes an overall phase factor to the wave function. Applying $\hat{S}_{{k}}(r_{k},\phi_{k})$ to the vacuum state then yields the wave function of the two-mode squeezed state:
\begin{equation}
|\psi\rangle= \hat{S}_{\vec{k}}(r_{k},\phi_{k})|0;0\rangle_{\vec{k},-\vec{k}}=\frac{1}{\cosh r_k}\sum_{n=0}^{\infty} (-1)^ne^{2in\phi_k}\tanh^nr_k|n;n\rangle_{\vec{k},-\vec{k}}. 
\label{squeezed state}
\end{equation}
With this wave function, we can now compute the evolution of the squeezed state, which is governed by the parameters \(r_k\) and \(\phi_k\). The derivation of the two-mode squeezed state in Eq.~\eqref{squeezed state} is model-independent, involving only the action of the squeeze operator on the vacuum. Model-specific information is encoded in the parameters \(r_k\) and \(\phi_k\), whose evolution is determined by the Hamiltonian. This will be confirmed in later sections. Furthermore, the two-mode quantum state \(|n;n\rangle_{\vec{k},-\vec{k}}\) is a Fock state, and its form is given by
\begin{equation}
|n;n\rangle_{\vec{k},-\vec{k}}=\bigg[\frac{1}{n!}(c_{\vec{k}}^\dagger c_{-\vec{k}}^\dagger)|0,0\rangle_{\vec{k},-\vec{k}}\bigg],
\label{fock state}
\end{equation}
thus, it automatically satisfies both orthogonality and normalization conditions. Here, we need to specify the initial Krylov operator $\mathcal{O}_0$ in our setup. As mentioned, the wave function is a two-mode squeezed state \eqref{squeezed state}; the initial operator must satisfy the Lanczos algorithm given in Eq.~\eqref{first two O}. Following Ref.~\cite{Adhikari:2022oxr}, we choose $|\mathcal{O}_0)=|0,0\rangle_{\vec{k},-\vec{k}}$ as our initial operator. This choice naturally fulfills the requirements of the Lanczos algorithm, which will be verified in Sec.~\ref{section op krylov complexity}.

\subsection{The evolution of $r_k(\eta)$ and $\phi_k(\eta)$}
\label{standard case}
 Numerical simulations of $\phi_k$ and $r_k$ require their evolution equations, which are derived from the Hamiltonian within the two-mode squeezed state formalism. This work focuses on evaluating the impact of the potential on Krylov complexity. The conventional Mukhanov-Sasaki variable is not ideal for this purpose, as it encapsulates the combined effect of the potential and other factors. To explicitly track the potential's role, we instead perform a direct second-order perturbation of the total action.
\begin{equation}
S=\frac{1}{2} \int d\eta d^{3}x\Bigl [f'^2-(\partial _{i}f)^2+(\frac{a'}{a})^{2}f^{2}-2f'f\frac{a'}{a}-a^2V_{,\phi \phi}f^2\Bigr ]
\label{action of standard case}
\end{equation}
where $f(\eta, \vec{x})$ is the inflaton perturbation defined via $\phi(\eta,\vec{x})=\bar{\phi}(\eta)+\frac{f(\eta, \vec{x})}{a(\eta)}$, with $\frac{f(\eta, \vec{x})}{a(\eta)} \ll 1$ in Planck units and this perturbation theory will be violated in strong gravitational system. In the current investigations, we focus on the perturbation of inflaton, which is responsible for the formation of large-scale structure; for further details, we refer the reader to \cite{Baumann:2009ds}. Under the slow-roll condition during inflation, we neglect the potential term, i.e., we assume
$\left(\frac{a'}{a}\right)^{2}f^{2} - 2f'f\frac{a'}{a} \gg a^2 V_{,\phi\phi} f^2 .
$
The Hamiltonian is then constructed as $H = \int d^{3}x \, d\eta \, (\pi f' - \mathcal{L})$, where $\mathcal{L}$ denotes the Lagrangian and $\pi$ is the conjugate momentum defined by
\begin{equation}
\pi(\eta,\vec{x})=\frac{\delta L}{\delta f'(\eta,\Vec{x})}=f'-\frac{a'}{a}f. 
\label{conjugate momentum}
\end{equation}
According to Eq. \eqref{action of standard case} and Eq. \eqref{conjugate momentum}, we obtain
\begin{equation}
H=\frac{1}{2}\int d^{3}xd\eta [\pi^{2}+(\partial_{i}f)^{2}+\frac{a'}{a}(\pi f+f\pi)+a^{2}V_{,\phi \phi}f^{2}]. 
\label{Hamiltonian of pi f}
\end{equation}
Using the Fourier decomposition of operator $\hat{f}(\eta,\Vec{x})$ and $\hat{\pi}(\eta,\Vec{x})$, 
\begin{equation}
\hat{f}(\eta,\Vec{x})=\int\frac{d^{3}k}{(2\pi)^{3/2}}\sqrt{\frac{1}{2k}}(\hat{c}_{\Vec{k}}^{\dagger}f_{\Vec{k}}^*(\eta)e^{-i\Vec{k\cdot}\Vec{x}}+\hat{c}_{\Vec{k}}f_{\Vec{k}}e^{i\Vec{k\cdot}\Vec{x}}),
\label{operator of field}
\end{equation}
\begin{equation}
\hat{\pi}(\eta,\Vec{x})=i\int\frac{d^{3}k}{(2\pi)^{3/2}}\sqrt{\frac{k}{2}}(\hat{c}_{\Vec{k}}^{\dagger}u_{\Vec{k}}^*(\eta)e^{-i\Vec{k\cdot}\Vec{x}}-\hat{c}_{\Vec{k}}u_{\Vec{k}}e^{i\Vec{k\cdot}\Vec{x}}). 
\label{operator of momentum}
\end{equation}
Substituting equations \eqref{operator of field} and \eqref{operator of momentum} into equation \eqref{Hamiltonian of pi f} to obtain the Hamiltonian operator, 
\begin{equation}
\hat{H}_k=(k+\frac{a^{2}V_{,\phi\phi}}{2k})(\hat{c}_{\Vec{k}}\hat{c}_{\Vec{k}}^{\dagger}+\hat{c}_{-\Vec{k}}^{\dagger}\hat{c}_{\Vec{-k}})+(\frac{a^{2}V_{,\phi\phi}}{2k}+i\frac{a'}{a})\hat{c}_{-\Vec{k}}^{\dagger}\hat{c}_{\Vec{k}}^{\dagger}+(\frac{a^{2}V_{,\phi\phi}}{2k}-i\frac{a'}{a})\hat{c}_{-\Vec{k}}\hat{c}_{\Vec{k}}. 
\label{Hamiltonian of standard case}
\end{equation}
Based on the Hamiltonian operator and Schr$\Ddot{o}$dinger equation
\begin{equation}
i\frac{d}{d\eta}|\psi\rangle=\hat{H}_k|\psi\rangle,
\label{Schrodinger equation}
\end{equation}
we obtain the evolution of equation of $r_k$ and $\phi_k$
\begin{equation}
\begin{split}
& r'_k=\frac{a^{2}V_{,\phi\phi}}{2k}\sin{2\phi _k}-\frac{a'}{a}\cos{2\phi _k}, \\&\phi _k^{'}=-k-\frac{a^{2}V_{,\phi \phi}}{2k}+(\frac{a'}{a}\sin{2\phi _k}+\frac{a^{2}V_{,\phi\phi}}{2k}\cos{2\phi _k})\coth{2r_k}.
\end{split}
\label{evolution in standard case}
\end{equation}
In inflation, Eq. \eqref{evolution in standard case} will nicely recover into the case of 
\begin{equation}
\begin{split}
& r_{k}^{'}=-\frac{a'}{a}\cos(2\phi_{k}),
\\& \phi_{k}^{'}=-k+ \frac{a'}{a}\sin(2\phi_{k})\coth(2r_{k}), 
\end{split}
\label{evolution in slow-roll}
\end{equation}
where it is consistent with \cite{Bhattacharyya:2020rpy} for $\frac{z'}{z}=\frac{a'}{a}$, since $\epsilon$ is constant. Directly solving Eq. \eqref{evolution in standard case} numerically is very difficult. Therefore, we introduce the variable transformation $y=\log_{10}a$ and solve Eq. \eqref{evolution in standard case} numerically in terms of $y$, 
\begin{equation}
\begin{split}
\frac{dr_k}{dy}=&\frac{\ln(10)10^{\frac{y}{2}(5+3\omega)}V_{,\phi\phi}}{2kH_0}\sin{2\phi_k}-\ln(10)\cos{2\phi_k}\\ \frac{d\phi_k}{dy}=&-\frac{\ln(10)10^{\frac{y}{2}(1+3\omega)}k}{H_0}-\frac{\ln(10)10^{\frac{y}{2}(5+3\omega)}V_{,\phi\phi}}{2kH_0}+\Bigl (\ln(10)\sin{2\phi_k}\\&+\frac{\ln(10)10^{\frac{y}{2}(5+3\omega)}V_{,\phi\phi}}{2kH_0}\cos{2\phi_k}\Bigr )\coth{2r_k} 
\end{split}
\label{numerical solution in standard case}
\end{equation}
where we have utilized Eq. \eqref{relation eta and a} and $\frac{a'}{a}=H_{0}a^{-\frac{1}{2}(1+3\omega)}$.

\begin{figure}
	\centering
	\includegraphics[width=.4\textwidth]{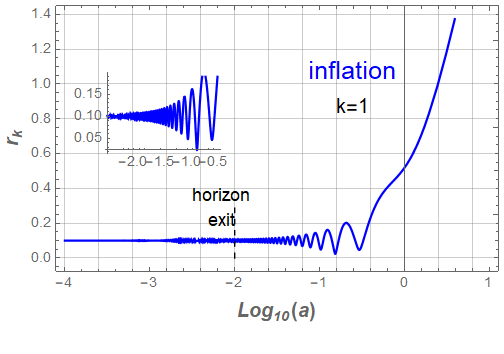}\\
	\qquad
	\includegraphics[width=.4\textwidth]{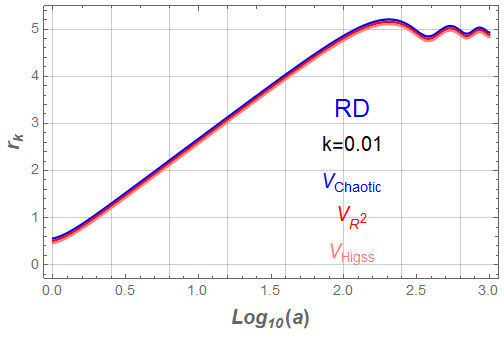}
	\qquad
	\includegraphics[width=.4\textwidth]{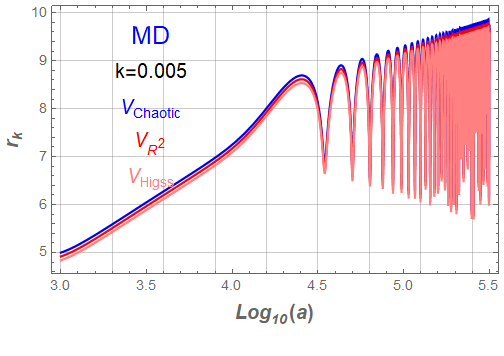}
	\caption{The numerical solution of $r_{k}$ in terms of $\log_{10}a$ for three different periods (inflation, RD, and MD), where we set $H_{0}=1$, $k=1$ in inflation, $k=0.01$ for RD, $k=0.005$ for MD. The effect of potential is always ignored in slow-roll inflation. But in RD and MD periods, the impact of potential is taken into account.}
	\label{fig: s rk}
\end{figure}
Before analyzing the figure for \( r_k \), it is important to note that the Krylov complexity depends solely on this quantity. Therefore, in Fig.~\ref{fig: s rk}, we show only the numerical results for \( r_k \) based on Eq.~\eqref{numerical solution in standard case}. For simplicity, we set \( H_0 = 1 \) and examine different values of the comoving wavenumber \( k \) across various cosmological epochs.

The first panel of Fig.~\ref{fig: s rk} illustrates the evolution of \( r_k \) during inflation. Prior to horizon exit, \( r_k \) exhibits oscillatory behavior, after which it grows linearly, consistent with the standard result reported in \cite{Li:2024kfm}. Following the transition to the RD era, \( r_k \) rises and then saturates, fluctuating around a nearly constant value. Notably, different inflationary potentials yield nearly identical trends for \( r_k \), indicating that the specific form of the potential does not significantly influence its evolution.

A similar trend is observed during the MD era, where the behavior of \( r_k \) closely resembles that in the RD era. However, we find that the value of the wavenumber \( k \) has a pronounced effect: as \( k \) decreases, \( r_k \) displays increasingly prominent oscillations. We note that presenting results in RD and MD, the contribution of potential is taken into account since the violation of slow roll condtions.

In this section, we have analyzed the evolution of \( r_k \) for different inflationary potentials within the single-field inflationary model. Our results indicate that \( r_k \) grows during the inflationary epoch. In both the RD and MD eras, \( r_k \) exhibits a similar qualitative behavior, saturating and then oscillating around an approximately constant value. Numerical analysis further shows that various inflationary models yield qualitatively similar trends for \( r_k \). Additionally, we find that both the scale factor and the comoving wavenumber significantly influence the evolution of \( r_k \). These features are summarized visually in Fig.~\ref{fig: s rk}.

\section{Krylov complexity in closed-system methodology}
\label{section Krylov complecity}
In this section, we conduct numerical simulations of Krylov complexity using the closed-system methodology \cite{Liu:2021nzx,Li:2021kfq,Li:2023ekd}. This framework encompasses both the Krylov complexity and the Krylov entropy. The Krylov complexity is determined by the coefficients $r_k$ and the phases $\phi_k$, which are in turn governed by the Hamiltonian operator given in Eq. \eqref{Hamiltonian of standard case}.

\subsection{Krylov complexity}

To obtain the Krylov complexity, the key quantity is the wave function. Specifically, the two-mode squeezed state wave function is expressed in the Krylov basis as follows,
\begin{equation}
\mathcal{O}(\eta)=\sum_{n=0}^{\infty} (i)^{n}\phi_{n}(\eta)|\mathcal{O}_{n})= \frac{1}{\cosh r_k}\sum_{n=0}^{\infty} (-1)^ne^{2in\phi_k}\tanh^nr_k|n;n\rangle_{\vec{k},-\vec{k}},
\label{two mode squeezed state1}
\end{equation}
where $\mathcal{O}(\eta)$ is the operator of wave function whose original definition can be found in Eq. \eqref{O with O_n}. 
Then, we find the explicit wave function $\phi_n$
\begin{equation}
\phi _{n}=\frac{1}{\cosh{r_k}}(-e^{2i\phi _k}\tanh{r_k})^{n}. 
\label{corresponding operator wave function}
\end{equation}
We define the Krylov compplexity in terms of $\phi_n$
\begin{equation}
K=\sum_{n} n\left |\phi _{n}\right |^{2}\\ =\cosh^{2}r_k\tanh^{2}r_k=\sinh^2 r_k. 
\label{Krylov complexity}
\end{equation}
This result explicitly demonstrates the independence of the Krylov complexity from $\phi_k$; hence, only $r_k$ is plotted.

\begin{figure}
	\centering
	\includegraphics[width=0.4\linewidth]{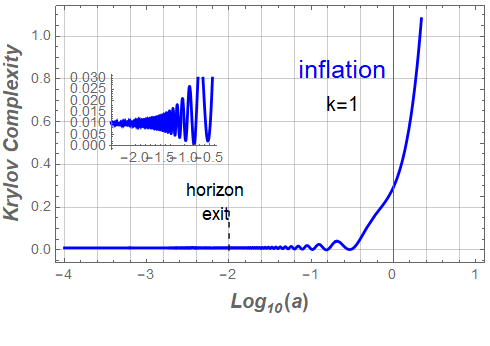}\\
	\qquad
	\includegraphics[width=0.4\linewidth]{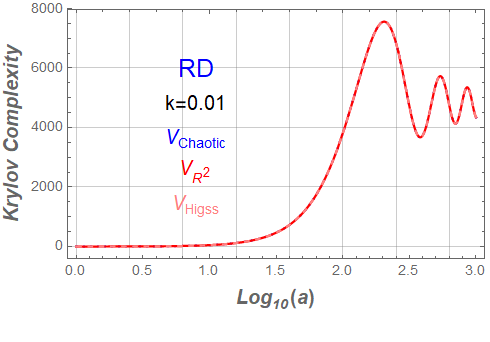}
	\qquad
	\includegraphics[width=0.4\linewidth]{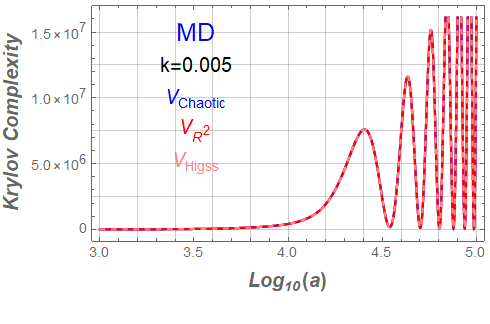}
	\caption{The numerical of Krylov complexity in terms of $\log_{10}a$ for three different periods (inflation, RD, and MD), where we set $H_{0}=1$, $k=1$ in inflation, $k=0.01$ at RD and $k=0.005$ at MD for simplicity.}
	\label{fig: s KC}
\end{figure}

Based on Eq. \eqref{Krylov complexity}, we show its numerical evolution in Fig. \ref{fig: s KC}, which depicts the behavior of the Krylov complexity throughout the inflationary, RD, and MD eras. Fig. \ref{fig: s KC} clearly reveals a significant growth in the Krylov complexity during the inflationary phase, a result consistent with the modified dispersion case studied in \cite{Li:2024kfm}. This finding is also in agreement with related work exploring Krylov complexity under exponential (de Sitter) expansion corresponding to inflation \cite{Adhikari:2022oxr}. We note that Ref. \cite{Adhikari:2022oxr} identifies the Krylov complexity with the average particle number, which scales proportionally with volume. However, we emphasize that the approach adopted in \cite{Adhikari:2022oxr} is specifically applicable to a closed system.

During the RD and MD periods, the Krylov complexity is observed to increase toward certain constant values, around which it subsequently fluctuates. Furthermore, different inflationary models exhibit broadly similar trends in the evolution of Krylov complexity across the respective epochs. Our results are consistent with the analysis in \cite{Barbon:2019wsy}, which found that in a closed system the Krylov complexity saturates to approximate constant values even after the scrambling time. The scrambling time itself will be analyzed in the open-system framework, specifically in relation to the Lanczos coefficients, the Lyapunov exponent, among other quantities.

\subsection{Krylov Entropy }
\label{section Krylov entropy}
\begin{figure}
	\centering
	\includegraphics[width=.4\textwidth]{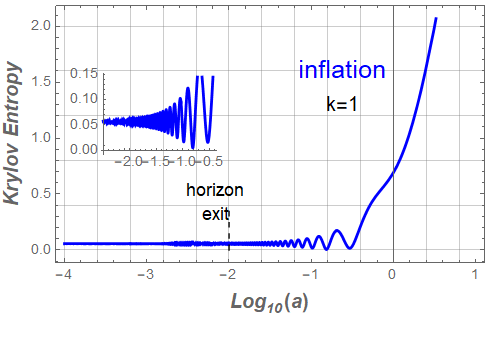}\\
	\qquad
	\includegraphics[width=.4\textwidth]{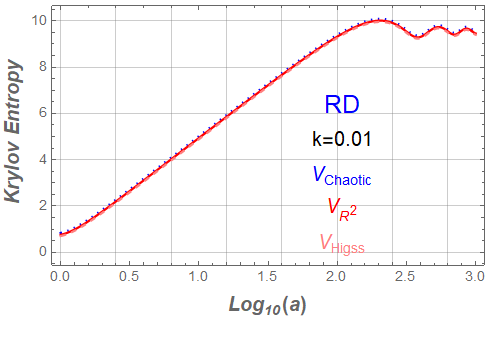}
	\qquad
	\includegraphics[width=.4\textwidth]{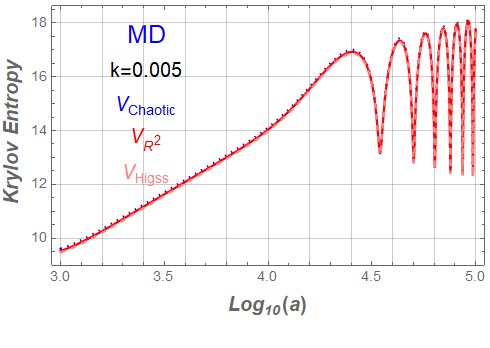}
	\caption{The numerical of Krylov entropy in terms of $\log_{10}a$ for three different periods (inflation, RD, and MD), where we set $H_{0}=1$, $k=1$ in inflation, $k=0.01$ at RD and $k=0.005$ at MD for simplicity.}
	\label{fig: s KE}
\end{figure}

In quantum systems, entropy is defined as a measure of a system's disorder. The Krylov entropy (K-entropy) serves to quantify the degree of disorder in curvature perturbations. Following \cite{Barbon:2019wsy}, the K-entropy is defined as follows,
\begin{equation}
\begin{split}
S_K&=-\sum_{n=0}^{\infty} \left|\phi_n\right|^{2}\ln\left|\phi_n\right|^{2} \\&=\cosh^{2}r_k\ln(\cosh^{2}r_k)-\sinh^{2}r_k\ln(\sinh^{2}r_k)
\end{split}
\label{Krylov entropy}
\end{equation}
where $\phi_{n}$ is the wave function \eqref{squeezed state}. Once given the analytic formula of K-entropy, we numerically simulate it in various epochs as shown in Fig. \ref{fig: s KE}.  

Fig. \ref{fig: s KE} illustrates the complete evolutionary profile of the K-entropy across the inflationary, RD, and MD eras, exhibiting a trend similar to that of the Krylov complexity. It can be observed that the curvature perturbations become progressively more disordered during inflation. In the subsequent RD and MD phases, the degree of disorder initially increases before saturating to approximately constant values; nonetheless, the overall trend remains one of growth, which is closely tied to the evolution of the scale factor. From another perspective, preheating, which generates particles during the MD era, also contributes to the observed increase in the K-entropy.

We now summarize the findings of Section \ref{section Krylov complecity}. In this section, we have conducted a comprehensive analysis of Krylov complexity in the early universe. Our numerical results clearly demonstrate that different inflationary potentials produce nearly identical patterns of evolution for both the Krylov complexity and the K-entropy. Specifically, the Krylov complexity grows during inflation and subsequently saturates to nearly constant values throughout the RD and MD eras. The K-entropy exhibits the same qualitative behavior. To characterize the chaotic features of the early universe more precisely, we must examine the Lanczos coefficients, which, in maximal chaos, play a role analogous to the Lyapunov exponent. These coefficients are identical in form for both the closed- and open-system approaches. In the following, we will employ the open-system methodology to investigate Krylov complexity.

\section{Krylov complexity with  open-system methodology}
\label{section op krylov complexity}
The preheating process during the MD and RD eras is a highly non-equilibrium phenomenon that generates particles \cite{Kofman:1997yn, Shtanov:1994ce}. Furthermore, Ref. \cite{Cheung:2007st} demonstrated that the evolution of the universe breaks time-reversal invariance, leading to a violation of energy conservation. For these reasons, the open-system methodology is regarded as more realistic than the closed-system approach. Several open-system methodologies exist for investigating Krylov complexity. Here, we will implement the generalization of the Lanczos algorithm to the Arnoldi iteration via the Heissenberg form \cite{Bhattacharjee:2022lzy,Bhattacharya:2022gbz,Liu:2022god,Nizami:2023dkf,Nandy:2024evd}. Let us review this Arnoldi iteration step by step. We begin by recalling the general Lindblad master equation \cite{Lindblad:1975ef}:
\begin{equation}
\dot\rho=-i[H,\rho]+\sum_{k}[L_k\rho L_k^\dagger-\frac{1}{2}\{L_k^\dagger L_k,\rho\}],
\label{general Lindblad}
\end{equation}
where $\rho$ is the density matrix, $H$ is the Hamiltonian, and $L_k^\dagger$ denotes the jump operator (which in our context corresponds to the annihilation operator). Following [81], we consider the open-system dynamics in an infinite-temperature environment. Specifically, we identify the system state $\rho$ with the maximally mixed state (infinite temperature state) $\rho_\infty$ due to the extremal high temperature for the early universe. This assumption defines the inner product for the operator space as $(A, B) = \text{Tr}(\rho_\infty A^\dagger B)$, which is central to our Krylov complexity calculations. Under this approximation, we obtain
\begin{equation}
\sum_{k}[L_k\rho_\infty L_k^\dagger-\frac{1}{2}\{L_k^\dagger L_k,\rho_\infty\}]=0,
\label{assumption for rho}
\end{equation} 
In this stationary state, it is reasonable to assume that the operator dynamics is governed by
\begin{equation}
\mathcal{\hat O}(t)=\exp(i\mathcal{L}t)\mathcal{\hat O},
\label{dynamic of operator}
\end{equation}
where the Heisenberg-picture Lindbladian $\mathcal{L}$ is given by \cite{Bhattacharya:2022gbz}, 
\begin{equation}
\mathcal{L}=\mathcal{L}_H+\mathcal{L}_D,~~\mathcal{L}_H\mathcal{\hat O}=[H,\mathcal{\hat O}],~~\mathcal{L}_D=	\sum_{k}[L_k\mathcal{\hat O} L_k^\dagger-\frac{1}{2}\{L_k^\dagger L_k,\mathcal{\hat O}\}]
\end{equation}
where $\mathcal{L}_H$ governs the Hermitian dynamics, $\mathcal{L}_D$ denotes the dissipative part, and $\mathcal{L}$ behaves as a superoperator on the operator space. The inner product is defined as $(A|B) := \mathrm{Tr}[\rho_\infty A^\dagger B]$, which explicitly connects the operator space structure to the Krylov basis, 
\begin{equation}
\rm span(\mathcal{O}_0,...,\mathcal{O}_n)=span (\mathcal{V}_0,...,\mathcal{V}_n), 
\label{span of krylov basis}
\end{equation}
where $\mathcal{V}_n$ denotes the original Krylov basis, which is generated via the Arnoldi iteration. In this work, we follow the standard construction, beginning with an initial normalized operator $\mathcal{O}_0 \propto \mathcal{O}$. Subsequent Krylov basis vectors are then generated by repeatedly applying the Liouvillian superoperator and performing orthonormalization at each step.
To compute the physical quantity, an appropriate basis is required. Following \cite{Bhattacharya:2022gbz}, we adopt an orthonormal basis $\{\mathcal{O}_0, \ldots, \mathcal{O}_n, \ldots\}$ generated by the open-system Lindbladian: 
\begin{equation}
\rm span(\mathcal{O}_0,...,\mathcal{O}_n)=span (\mathcal{V}_0,...,\mathcal{V}_n), 
\label{span of krylov basis}
\end{equation}
where $\mathcal{V}_n$ is the original Krylov basis, and we generate this basis by using the Arnoldi iteration, which is started with an initial normalized vector $\mathcal{O}_0 \propto \mathcal{O}$. Thereafter, we construct the subsequent Krylov basis vectors by repeating the following steps:
\begin{equation}
|\mathcal{U}_k)=\mathcal{L}| \mathcal{O}_{k-1}).
\label{recursion relation}
\end{equation}
From $j=0$ to $n-1$, we perfom the following iterations: $1.~ h_{j,k-1}=(\mathcal{O}_j|\mathcal{U}_k)$, $2.~|\tilde{\mathcal{U}}_k)=|\mathcal{U}_k)-\sum_{j=0}^{k-1}h_{j,k-1}|\mathcal{O}_j)$, $3.~h_{k,k-1}=\sqrt{(\tilde{U}_k | \tilde{\mathcal{U}}_k)}$, it will stop as if $h_{k,k-1}=0$, otherwise, it is defined as follows,
\begin{equation}
|\mathcal{O}_k)=\frac{|\tilde{\mathcal{U}}_k)}{h_{k,k-1}}.
\label{final recursion}
\end{equation}
Finally, the Lindbladian can transform into an upper Heissenberg form in the span Krylov
basis: 
\begin{equation}
\mathcal{L}=
\begin{pmatrix}
h_{0,0} & h_{0,1} & h_{0,2} & \cdots & \cdots & h_{0,n} \\
h_{1,0} & h_{1,1} & h_{1,2} & \cdots & \cdots & h_{1,n} \\
0 & h_{2,1} & h_{2,2} & h_{2,3} & \cdots & \cdots \\
\vdots & 0 & h_{3,2} & \cdots & \cdots & \cdots \\
0 & \vdots & 0 & \vdots & \ddots & h_{n-1,n} \\
0 & 0 & \vdots & 0 & h_{n,n-1} & h_{n,n}
\end{pmatrix}
\label{matrix of L}
\end{equation}
where $h_{m,n}=(\mathcal{O}_m|\mathcal{L}|\mathcal{O}_n)$. As $\mathcal{L}$ is a Hermitian superoperator, the Arnoldi iteration will reduce into the Lanczos algorithm.
Following conventions of \cite{Li:2024kfm}, we represent the generalized Lanczos algorithm as
\begin{equation}
\mathcal{L}|\mathcal{O}_n)=-ic_n|\mathcal{O}_n)+b_{n+1}|\mathcal{O}_{n+1})+b_n|\mathcal{O}_{n-1}),
\label{eq L in K-basis}
\end{equation}
where the coefficient \(c_n\) encodes the information of the open quantum system dynamics for a given Hamiltonian, corresponding to the diagonal elements \(h_{n,n}\) in Eq. \eqref{matrix of L}. Meanwhile, the Lanczos coefficients \(b_n\), associated with the off-diagonal elements \(h_{n,n-1}\) in Eq. \eqref{matrix of L}, characterize the chaotic behavior of the dynamical system. The orthogonal basis \(\mathcal{O}_n = |n,n\rangle_{-\vec{k},\vec{k}}\) spans the Krylov space defined by Eq. \eqref{span of krylov basis}. The Liouvillian superoperator \(\mathcal{L}\) is equivalent to a Hamiltonian operator that admits a representation in terms of creation and annihilation operators.

\subsection{Lanczos coefficient and dissipation coefficient}
 
Based on the foregoing analysis, we separate the total Hamiltonian (the Lindbladian) into two constituent parts:
\begin{equation}
\mathcal{L}_{o}=\mathcal{H}_{k}=\mathcal{H}_{close}+\mathcal{H}_{open}
\end{equation}
 In light of \eqref{eq L in K-basis}, we easily obtain the open part (generating $c_n$) of Hamiltonian, 
\begin{equation}
\mathcal{H}_{open}=
(k+\frac{a^{2}V_{,\phi\phi}}{2k})(\hat{c}_{-\vec{k}}^{\dagger}\hat{c}_{-\vec{k}}+\hat{c}_{\vec{k}}\hat{c}_{\vec{k}}^{\dagger})
\label{eq H open}
\end{equation}
and the close part (generating $b_n$ and $b_{n-1}$) is
\begin{equation}
\mathcal{H}_{close}=
(\frac{a^{2}V_{,\phi\phi}}{2k}+i\frac{a'}{a})\hat{c}_{\vec{k}}^{\dagger}\hat{c}_{-\vec{k}}^{\dagger}+(\frac{a^{2}V_{,\phi\phi}}{2k}-i\frac{a'}{a})\hat{c}_{\vec{k}}\hat{c}_{-\vec{k}}
\label{eq H close}
\end{equation}
where $|\mathcal{O}_{n})=\Bigl | n_{\vec{k}};n_{-\vec{k}}\Bigr \rangle$. Specifically, the jump operators $L_k$ appearing in the dissipative term $L_D$ are identified with the annihilation and creation operators in our work, i.e., $L_{k} = c_k$ and $L_{-k} = c_{-k}^\dagger$. By substituting Eq. \eqref{eq H open} and Eq. \eqref{eq H close} into Eq. \eqref{eq L in K-basis}, we obtain the equations for $c_{n}$ and $b_{n}$ as
\begin{equation}
c_{n}=
i(2n+1)(k+\frac{a^{2}V_{,\phi\phi}}{2k}).
\label{eq cn}
\end{equation}
\begin{equation}
b_{n}=
n\sqrt{[(\frac{a^{2}V_{,\phi\phi}}{2k})^{2}+(\frac{a'}{a})^{2}]}.
\label{eq bn}
\end{equation}
Here, we provide further discussion regarding the coefficients $b_n$ and $c_n$. In Ref.~\cite{Bhattacharjee:2022lzy}, the coefficient denoted by $a_n$ corresponds to our $c_n$, and it is also established that $c_n$ is associated with the dissipation coefficient. Meanwhile, we observe that $b_n$ remains identical for both the closed and open systems, as it is determined solely by the Hamiltonian.

In Ref.~\cite{Parker:2018yvk}, it is clearly established that the dynamical system is an infinite, chaotic, many-body system as \( b_n = \alpha n + \gamma \), where \(\alpha\) and \(\gamma\) are parameters that depend on the specific model. In our case, since \( b_n \propto n \), the early universe can be regarded as an ideal chaotic system.

\subsubsection{$b_n$ and $\mu_2$}
 In this work, we adopt the notation of Ref.~\cite{Li:2024kfm} to perform numerical simulations of the coefficients $b_n$ and $\mu_2$. We begin by examining the properties of $c_n$ within our framework. Recall that Eq.~\eqref{O in SE} (which is essentially the Schr\"odinger equation) involves only the Lanczos coefficients. The parameter $c_n$ will subsequently appear in the following modified equation:
\begin{equation}
\partial_\eta\phi+2b_n\partial_n\phi+\tilde{c}_n\phi=0,
\label{equ of cn}
\end{equation}
Here, we have defined \(\tilde{c}_n = -i c_n\). In taking the continuous limit, we have employed the approximations \(b_{n+1} \approx b_n\) and \(\phi_{n+1} - \phi_{n-1} \approx 2 \partial_n \phi\). The coefficients \(b_n\) and \(c_n\) are defined in Eq.~\eqref{eq bn} and Eq.~\eqref{eq cn}, respectively. Their general expressions are given by
\begin{equation}
b_{n}^{2}=|1-\mu_{1}^{2}|n(n-1+\beta),\ \ c_{n}=i\mu_{2}(2n+\beta)
\label{eq bn and cn}
\end{equation}
where $\mu_1$, $\mu_2$, $\beta$ are determined by various models and $\mu_2$ plays a role of dissipative coefficient as shown in Ref. \cite{Li:2024kfm}. In our work, one can explicitly find $\beta=1$ and combine Eqs. \eqref{eq cn}, \eqref{eq bn}, and \eqref{eq bn and cn}, we find that the parameters $\mu_{1}$ and $\mu_{2}$ have the following forms,
\begin{equation}
|1-\mu_{1}^{2}|=
(\frac{a^{2}V_{,\phi\phi}}{2k})^{2}+(\frac{a'}{a})^{2}
,\mu_{2}=
k+\frac{a^{2}V_{,\phi\phi}}{2k}. 
\label{eq mu1 mu2}
\end{equation}
It should be noted that during inflation, the parameter \(\mu_2\) is primarily determined by the wavenumber \(k\) due to the slow-roll conditions. Upon entering the RD and MD eras, which is taking into account the effective mass, we observe that \(\mu_2\) becomes smaller than unity, because the effective mass associated with these three types of potentials is much smaller than the momentum \(k\).

\begin{figure}
	\centering
	\includegraphics[width=.4\textwidth]{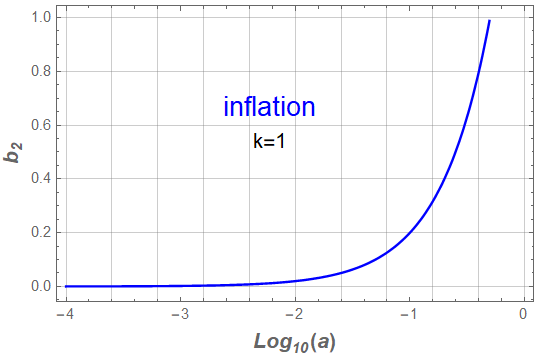}\\
	\qquad
	\includegraphics[width=.4\textwidth]{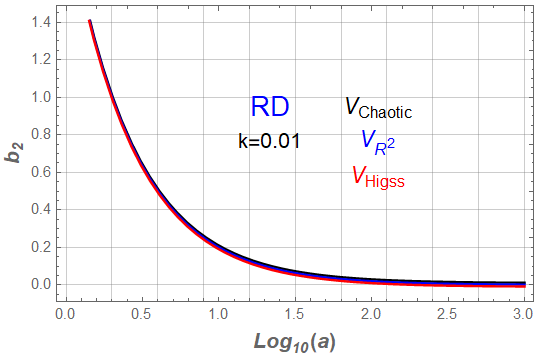}
	\qquad
	\includegraphics[width=.4\textwidth]{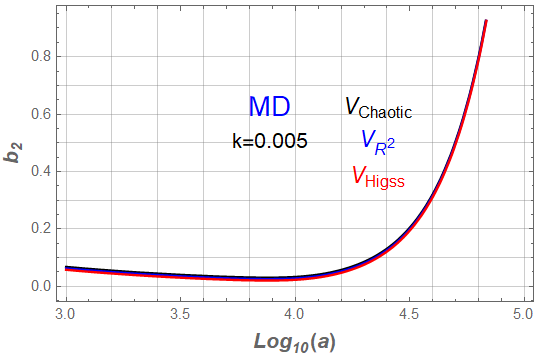}
	\caption{The plot of $b_n=\lambda$ with $n=2$.The numerical of $b_{n}$ in terms of $\log_{10}a$ for three different periods (inflation, RD, and MD), where we set $H_{0}=1$, $k=1$ in inflation, $k=0.01$ at RD and $k=0.005$ at MD for simplicity.}
	\label{fig: op s bn}
\end{figure}

We first discuss the Lanczos coefficient \(b_n\). Figure~\ref{fig: op s bn} clearly illustrates the three distinct behaviors of \(b_n\) during inflation, RD, and MD, where the parameters \(k\) and \(H_0\) are set to the same values as in the previous plots. As established in Ref.~\cite{Bhattacharya:2022gbz}, the Lyapunov exponent \(\lambda\) is related to \(b_n\) by \(\lambda = 2\alpha\), given the linear form \(b_n = \alpha n + \delta\) (with \(\delta = 0\) in our case). This explicitly shows that \(\lambda = 2\alpha\), or equivalently \(\lambda = b_n\) for \(n = 2\), as will be demonstrated in Figure~\ref{fig: op s bn}. Physically, \(\lambda\) characterizes the chaotic nature of the dynamical system. Consequently, in our framework, \(b_n\) also reflects the chaotic behavior of the early universe. 

Figure \ref{fig: op s bn} illustrates the evolution of the coefficient \( b_n \) during inflation, RD, and MD. Based on the analysis of \( b_n \), the degree of chaos increases due to the exponential expansion of the background spacetime. As the energy scale decreases, \( b_n \) decreases correspondingly with the temperature. During MD, particles are produced via the preheating mechanism known as parametric resonance, leading to an increase in \( b_n \).

\begin{figure}
	\centering
	\includegraphics[width=.4\textwidth]{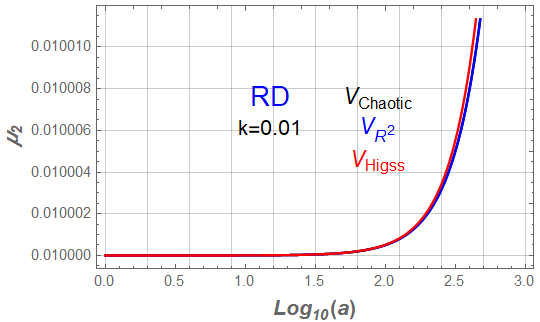}
	\qquad
	\includegraphics[width=.4\textwidth]{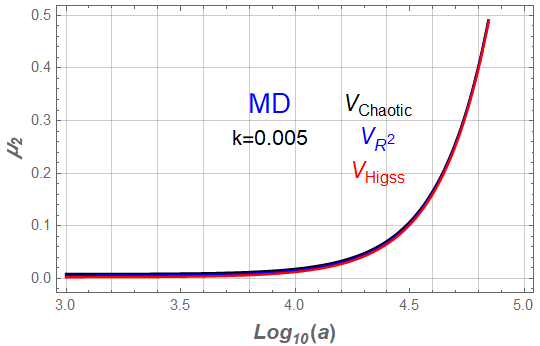}
	\caption{The numerical of dissipative coefficient $\mu_{2}$ in terms of $\log_{10}a$ for RD and MD, where we set $H_{0}=1$, $k=1$ in inflation, $k=0.01$ at RD and $k=0.005$ at MD for simplicity. The value of $\mu_2$ indicates the RD and MD are weak dissipative systems.}
	\label{fig: op s u2}
\end{figure}

Finally, we analyze the dissipation coefficient \( \mu_2 \), which is related to \( c_n \), thereby confirming previous results from Ref.~\cite{Li:2024kfm}. Figure \ref{fig: op s u2} clearly shows the evolution of \( \mu_2 \) in RD and MD. The value of \( \mu_2 \) during inflation is not shown in Fig.~\ref{fig: op s u2} because \( \mu_2 = k = 1 \) for the chosen wavenumber \( k \) in the inflationary scenario. As asserted in Ref.~\cite{Li:2024kfm}, inflation constitutes a strong dissipative system with \( \mu_2 \geq 1 \), while RD and MD are weak dissipative systems with \( \mu_2 \ll 1 \), a result explicitly corroborated by Fig.~\ref{fig: op s u2}.

To summarize the chaotic features of the early universe: First, chaos is enhanced by the exponential expansion. Then, as the energy scale diminishes, chaos decreases. Finally, the generation of particles during MD through the preheating process increases chaos once again. 

\subsection{General discussion of $\phi_n$}
\label{general discussion of phi}
Before investigating the Krylov complexity, we discuss the general properties of $\phi_n$ since the definition of Krylov complexity is explicitly represented by $C_K=\sum_{n=1}^{+\infty}n|\phi_n|^2$. We will follow the general discussion of $\phi_n$ via Refs. \cite{Li:2024kfm, Bhattacharjee:2022lzy}. Eq. \eqref{equ of cn} is rewritten by
\begin{equation}
\partial_\eta\phi+n(\chi\mu \phi+2\alpha\partial_n\phi)=0,
\label{eq of cn1}
\end{equation}
where we have defined $\tilde{c}_n=n\chi\mu$ and $b_n=n\alpha$. As $\eta$ approaches infinity, the solution of Eq. \eqref{eq of cn1} has the form $\phi(n)\propto e^{-n\xi}$ with $\xi=\frac{2\alpha}{\mu\chi}$, where this form will appear in RD or MD due to their weak dissipative nature. In the weak dissipative region, the corresponding Krylov complexity behaves as follows: it increases exponentially at the onset, with the average position given by \(C_k \propto e^{2\alpha\eta}\) before reaching \(\eta_*\). After this point, the Krylov complexity saturates to a constant value \(\xi\) when \(\eta > \eta_*\), where \(\eta_* \sim \frac{1}{2\alpha} \ln{\frac{2\alpha}{\chi\mu}}\) dubbed as the scrambling time. It is evident that $\eta_*$ determines the range of exponential growth of Krylov complexity, determined by the dynamical system itself. It is straightforward to demonstrate that \(\exp(2a\eta_*) = \xi\). After some simple algebra, we obtain the formula $\xi$ in our case as follows,
\begin{equation}
\xi=\frac{2\sqrt{(\frac{a^2V_{,\phi\phi}}{2k})^2+(\frac{a'}{a})^2}}{(2+\frac{1}{n})(k+\frac{a^2V_{,\phi\phi}}{2k})}
\end{equation}
From this formula, we observe that $\xi$ is not strictly constant; it varies with respect to $a$ and $k$. Consequently, a significant change in the background will have a substantial impact on Krylov complexity. In this section, we examined the general properties of Krylov complexity without providing the precise formula for $\phi_n$. To fully capture the information concerning Krylov complexity within the framework of the open-system methodology, it is essential to have the exact wave function for $\phi_n$. 

\subsection{Wave function within open-system methodology}
Although we constructed the wave function for an open system in reference \cite{Li:2024kfm}, based on the framework established in \cite{Bhattacharjee:2022lzy},  this wave function cannot capture the complete information since it is the function of the conformal time $\eta$. To clearly illustrate this issue, we will first present our constructed wave function as follows:
\begin{equation}
\phi_{n}=\frac{\rm sech \eta}{1+\mu_{2}\tanh \eta}|1-\mu_{1}^{2}|^{\frac{n}{2}}(\frac{\tanh \eta}{1+\mu_{2}\tanh\eta})^{n}. 
\label{eq wave function of op}
\end{equation}
To perform the Taylor expansion around \(\mu_2\) to the leading order $(\mu_2=0)$, we obtain the expression 
\[
\phi_n \approx  \text{sech}(\eta) \, \tanh^n(\eta) + \mathcal{O}(\mu_2^n).
\] To relate with Eq, \eqref{corresponding operator wave function}, we can find one explicit correspondence between $r_k$ and $\eta$. To include the parameter of $\phi_k$, we will follow the guidelines provided in Appendix D of \cite{Parker:2018yvk} to rigourisly derive the so-called open two-mode squeezed state, which will accurately reproduce the wave function represented in \eqref{corresponding operator wave function}.

For deriving the two-mode squeezed state in open system, we first introduce the concept of orthogonal polynomial sequence (OPS) $\{P_n(x)\}_{n\geq 0}$, it satisfied with the following recurrence relation, 
\begin{equation}
P_{n+1}(x)=(x-\tilde{c}_{n})P_{n}(x)-b_{n}^{2}P_{n-1}(x),\ \ \ \ \mbox{for}\ \  n\geq 1
\label{recurrence relation}
\end{equation}
where \( x \) represents the Hamiltonian, \( \tilde{c}_n \) is associated with the dissipative coefficient, and \( b_n \) refers to the Lanczos coefficient. The sequence begins with the initial conditions \( P_{0}(x) = 1 \) and \( P_{1}(x) = x - c_{0} \). For an open system, we have \( P_{n}(x;\alpha,\eta) = \det(x - \mathcal{L}_{n}) \), where \( \mathcal{L}_n \) is the Hamiltonian for the \( n \)-th quantum state. This determinant encodes information about \( b_n \) as represented by the corresponding matrix. Based on the definition of \( L \), we express it in terms of \( x^n \) or \( e_n \), which will be used to derive the wave function using the open-system methodology.

To obtain the similar recurrence relation as \eqref{eq L in K-basis}, we introduce the the natural orthonormal basis $\{ e_{n}\}$, and meanwhile $P_n$ is represented in terms of $e_n$
\begin{equation}
P_{n}(x)=\Bigl (\prod_{k=1}^{n}b_{k}\Bigr )e_{n},\ \ \ \ \mbox{and}\ \ \ x^{n}=\mathcal{L}^{n}e_{0},
\label{definition of pn}
\end{equation}
where \( x \) is the parameter of \( P_n \), \( \mathcal{L} \) represents our Hamiltonian expressed in terms of creation and annihilation operators, and \( |e_n\rangle \equiv |\mathcal{O}_n\rangle = \Bigl| n_{\vec{k}}; n_{-\vec{k}} \Bigr\rangle \). By combining Equation \eqref{recurrence relation} with Equation \eqref{definition of pn}, we derive the following recurrence relation.
\begin{equation}
b_{n+1}e_{n+1}+b_n e_{n-1}=(x-\tilde{c}_n)e_n. 
\label{recurrence relation of en}
\end{equation} 
Here, $x$ functions as the Hamiltonian. Given the recurrence of $P_n$, it aligns with the second kind of Meixner polynomials, denoted as $M_{n}(x;\delta,\eta_1)$. Its generating function is
\begin{equation}
\sum_{n=0}^{\infty}M_{n}(x;\delta,\eta_1)\frac{t^{n}}{n!}=((1+\delta t)^{2}+t^{2})^{-\eta /2}\exp(x\arctan(\frac{t}{1+\delta t}))
\label{generation function of Pn}
\end{equation}
The inverse polynomial of $\{ M_{n}(x;\delta,\eta_1)\} _{n\geq 0}$ is $\{ Q_{n}(x;\delta,\eta_1)\} _{n\geq 0}$, which is defined as 
\begin{equation}
Q_{n}(x)=\sum_{k=0}^{n}\mu_{n,k}x^{k},\ \ \ \ \mbox{and} \ \ \ x^{n}=\sum_{k=0}^{n}\mu_{n,k}P_{k}(x).
\label{Qn and xn}
\end{equation}
where  $\mu_{n,k}$ refers to a triangular linear transform with matrix elements that relate \( x \) to \( P_n \). This formula demonstrates that the polynomials \( P_n \) is expressed in terms of \( x \). All three orthogonal polynomial sequences (OPS) mentioned are examples of Sheffer orthogonal polynomials, which are defined as a set of polynomials \( \{P_{n}(x)\}_{n \geq 0} \) characterized by a generating function,  
\begin{equation}
\sum_{n=0}^{\infty}P_{n}(x)\frac{t^{n}}{n!}=f(t)\exp(xg(t)). 
\label{gneration function of Pn1}
\end{equation}
And the inverse polynomials are give by
\begin{equation}
\sum_{n=0}^{\infty}Q_{n}(x)\frac{t^{n}}{n!}=\frac{1}{f(g^{\langle -1\rangle}(t))}\exp(xg^{\langle -1\rangle}(t)),
\label{gneration function of Qn}
\end{equation}
where $g^{\langle -1\rangle}(t)$ stands for the compositional inverse of $g(t)$, which have the following relationship
\begin{equation}
g\bigl (g^{\langle -1\rangle}(t)\bigr )=t. 
\label{g}
\end{equation}
Back to the Meixner polynomials of the second kind, we find that
\begin{equation}
f(t)=((1+\delta t)^2+t^{2})^{-\eta /2},\ \ \ \ \mbox{and}\ \ \ g(t)=\arctan(\frac{t}{1+\delta t}).
\label{f and g}
\end{equation}
Our target function is \( Q_n(x) \), and the rationale will be provided in the final derivations. To obtain \( Q_n(x) \), we need to know the formula for \( g^{\langle -1\rangle} \). The compositional inverse polynomial of \( g(t) \) is
\begin{equation}
t=g\bigl (g^{\langle -1\rangle}(t)\bigr )=\arctan(\frac{g^{\langle -1\rangle}(t)}{1+\delta g^{\langle -1\rangle}(t)}),
\label{t for g}
\end{equation}
Next,
\begin{equation}
\frac{g^{\langle -1\rangle}(t)}{1+\delta g^{\langle -1\rangle}(t)}=\tan(t)
\end{equation}
And after a few simple calculations,
\begin{equation}
g^{\langle -1\rangle}(t)=\frac{\tan(t)}{1-\delta \tan(t)},
\label{inverse of g}
\end{equation}
where $\delta$ is a parameter determined by various models. Combine Eqs. \eqref{inverse of g} and \eqref{gneration function of Qn}, we obtain the inverse polynomials of $\{ M_{n}(x;\delta,\eta_1)\} _{n\geq 0}$ as following, 
\begin{equation}
\sum_{n=0}^{\infty}Q_{n}(x)\frac{t^{n}}{n!}=\frac{\sec(t)^{\eta_1}}{(1-\delta \tan(t))^{\eta_1}}\exp(x\frac{\tan(t)}{1-\delta \tan(t)}).
\label{generation function of Qn1}
\end{equation}
After the changing the variable as $t=i\eta$ and $\delta=i\mu_2$, where $\eta$ (conformal time) and $\mu_2$ (dissipation coefficient) are our defined parameters. Making use of $\tan i\eta=i\tanh \eta$ and $\cos i\eta=\cosh \eta$,  the generating function of the inverse polynomials become
\begin{equation}
\sum_{n=0}^{\infty}Q_{n}(x)\frac{(i\eta)^{n}}{n!}=\frac{\rm sech(\eta)^{\eta_1}}{(1+\mu_2 \tanh(\eta))^{\eta_1}}\exp(x\frac{i\tanh(\eta)}{1+\mu_2\tanh(\eta)}).
\end{equation}
Based on Eqs. \eqref{definition of pn} and \eqref{Qn and xn}, we obtain 
\begin{equation}
(e_d| \mathcal{L}^n|e_0)=\mu_{n,d} \prod_{k=1}^{n}b_{k}.
\label{matrix}
\end{equation}
Furthermore, we have 
\begin{equation}
(e_{d}|e^{i\mathcal{L}\eta}|e_{0})=\prod_{k=1}^{d}b_{k}\sum_{n=0}^{\infty}\frac{(i\eta)^{n}}{n!}\mu_{n,d}.
\label{matrix element}
\end{equation}
In light of \eqref{Qn and xn}, we obtain 
\begin{equation}
\sum_{n=0}^{\infty}\frac{(i\eta)^{n}}{n!}\mu_{n,d}=\frac{\rm sech \eta}{1+\mu_2\tanh \eta}\frac{1}{d!}\frac{(i\tanh \eta)^{d}}{(1+\mu_2\tanh \eta)^{d}},
\end{equation}
Then we utilize $b_n=\alpha n =\sqrt{|1-\mu_1^2|}n$ in our case, we naturally get the following formula,
\begin{equation}
(e_{n}|e^{i\mathcal{L}\eta}|e_{0})=\frac{\rm sech \eta}{1+\mu_2\tanh \eta}|1-\mu_1^{2}|^{\frac{n}{2}}\frac{(i\tanh \eta)^{n}}{(1+\mu_2\tanh \eta)^{n}}
\end{equation}
thus, we obtain the wave function
\begin{equation*}
|\mathcal{O}(t))=e^{i\mathcal{L}\eta}|e_{0})=\frac{\rm sech \eta}{1+\mu_2\tanh \eta}\sum_{n=0}^{\infty}|1-\mu_1^{2}|^{\frac{n}{2}}\frac{(-1\exp(2i\phi_k(\eta))\tanh \eta)^{n}}{(1+\mu_2\tanh \eta)^{n}}|e_{n})
\end{equation*}
where 
\begin{equation*}
|\mathcal{O}_{n})=|e_{n})
\end{equation*}
In our calculations, we multiply a phase factor \( ie^{2i\phi_k(\eta)} \) which does not alter the wave function. Finally, we set \( r_k = \eta \), allowing us to denote the two-mode squeezed state within the open system as, 
\begin{equation}
\phi_{n}(\eta)=(-1)^ne^{2in\phi_k}\frac{\rm sech r_k}{1+\mu_2\tanh r_k}|1-\mu_1^{2}|^{\frac{n}{2}}\frac{(\tanh r_k)^{n}}{(1+\mu_2\tanh r_k)^{n}}.
\label{wave function of open system}
\end{equation}
This wave function constitutes the central result of this paper.
Its leading-order expansion of \(\phi_n\) in terms of \(\mu_2\) is consistent with the wave function presented in \eqref{corresponding operator wave function}.
The idea of performing a variable change is inspired by the construction of coherent states,
where a key element is the displacement operator, defined as
\[
D(\eta) = e^{i H t},
\]
with \(H\) being the corresponding Hamiltonian, and where the parameter \(t\) can be re-expressed in terms of \(r_k\) and \(\phi_k\).
As noted in Ref.~\cite{Caputa:2021sib}, a generalized coherent state can be derived by constructing a generalized displacement operator based on the appropriate group representation theory.

\subsection{The power of wave function \eqref{wave function of open system}}

We discuss the two-mode squeezed state in an open system, as described in equation \eqref{wave function of open system}. In contrast to the two-mode squeezed state derived from closed-system methedology, our wave function \eqref{wave function of open system} is more physically realistic, since it includes the contribution of open system via Hamiltonian.
Furthermore, although our derivation is obtained within the context of the early universe, this wave function is also widely applicable in fields such as quantum optics, quantum information, and condensed matter physics, where physical information is encoded in the parameters $\mu_1$ and $\mu_2$, which are determined by the specific system Hamiltonian.
Finally, we verify the validity of our results by requiring that the recurrence relation \eqref{recurrence relation}, which is equivalent to the Lanczos algorithm, be satisfied. This connection underscores the utility of the Lanczos algorithm, which offers a systematic framework for constructing the wave function directly from the Hamiltonian, even for open quantum systems.

\subsection{The evolution of $r_k(\eta)$ and $\phi_k(\eta)$ in open-system methodology}
Following the standard procedure, we derive the evolution equations for $r_k$ and $\phi_k$ with respect to conformal time $\eta$. The details of this derivation are provided in Appendix \ref{Appendix A}. The resulting evolution is summarized by the following equations:
\begin{equation}
\begin{split}
r_{k}'=&\ \frac{|1-\mu_{1}^{2}|^{\frac{1}{2}}\sinh 2r_{k}\bigl (\frac{a^{2}V_{,\phi\phi}}{2k}\sin 2\phi_{k}-\frac{a'}{a}\cos 2\phi_{k}\bigr )-\sinh 2r_{k}\mu_{2}'}{(\sinh 2r_{k}+2\mu_{2}\cosh^{2}r_{k})}\\
\phi_{k}' =&\ -\bigl (k+\frac{a^{2}V_{,\phi\phi}}{2k}\bigr )+\frac{1}{2}\bigl (\frac{a^{2}V_{,\phi\phi}}{2k}\cos 2\phi_{k}+\frac{a'}{a}\sin 2\phi_{k}\bigr )\\
&\ \Bigl [|1-\mu_{1}^{2}|^{\frac{1}{2}}\frac{\tanh r_{k}}{1+\mu_{2}\tanh r_{k}}+|1-\mu_{1}^{2}|^{-\frac{1}{2}}(\tanh^{-1}r_{k}+\mu_{2})\Bigr ],
\end{split}
\label{evolution in ete}
\end{equation}
where prime denotes the varying with respect to the conformal time $\eta$. We change into the variable $y=\log_{10}a$, the evolution equation of $r_k$ and $\phi_k$ will become as,
\begin{equation}
\begin{split}
\frac{H_{0}}{\ln 10}10^{-\frac{y}{2}(1+3\omega)}\frac{dr_{k}}{dy}=&\ \frac{|1-\mu_{1}^{2}|^{\frac{1}{2}}\sinh 2r_{k}\bigl (\frac{10^{2y}V_{,\phi\phi}}{2k}\sin 2\phi_{k}-H_{0}10^{-\frac{y}{2}(1+3\omega)}\cos 2\phi_{k}\bigr )-\sinh 2r_{k}\mu_{2}'}{(\sinh 2r_{k}+2\mu_{2}\cosh^{2}r_{k})}\\
\frac{H_{0}}{\ln 10}10^{-\frac{y}{2}(1+3\omega)}\frac{d\phi_{k}}{dy}=&\ -\bigl (k+\frac{10^{2y}V_{,\phi\phi}}{2k}\bigr )+\frac{1}{2}\bigl (\frac{10^{2y}V_{,\phi\phi}}{2k}\cos 2\phi_{k}+H_{0}10^{-\frac{y}{2}(1+3\omega)}\sin 2\phi_{k}\bigr )\\
&\ \Bigl [|1-\mu_{1}^{2}|^{\frac{1}{2}}\frac{\tanh r_{k}}{1+\mu_{2}\tanh r_{k}}+|1-\mu_{1}^{2}|^{-\frac{1}{2}}(\tanh^{-1}r_{k}+\mu_{2})\Bigr ]
\end{split}
\label{evolution in loga}
\end{equation}
where 
\begin{equation*}
\begin{split}
&\ \mu_{2}=k+\frac{10^{2y}V_{,\phi\phi}}{2k},\ \ \ \ \mu_{2}'=H_{0}10^{-\frac{y}{2}(1+3\omega)}\frac{10^{2y}V_{,\phi\phi}}{k}+\frac{10^{2y}V_{,\phi\phi}'}{2k}\\ 
&\ |1-\mu_{1}^{2}|=(\frac{10^{2y}V_{,\phi\phi}}{2k})^{2}+(H_{0}10^{-\frac{y}{2}(1+3\omega)})^{2}. 
\end{split}
\end{equation*}
During the calculations, we accounted for the background of \(\phi\) using conformal time \(\eta\) via preheating, as demonstrated by the various potentials in Eq. \eqref{eq vi phi phi}. Consequently, \(\mu_2\) and \(\mu_1\) also vary with respect to conformal time, since they include contributions from the potential and the scale factor \(a\).
\begin{figure}
	\centering
	\includegraphics[width=.4\textwidth]{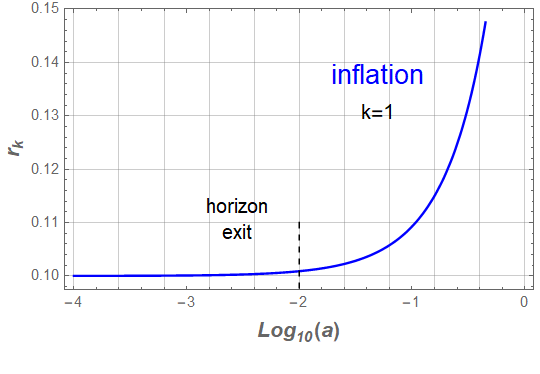}\\
	\qquad
	\includegraphics[width=.4\textwidth]{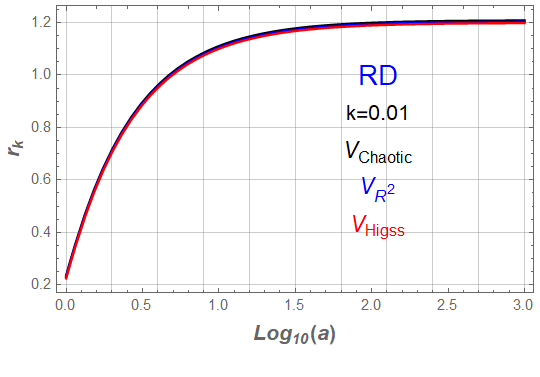}
	\qquad
	\includegraphics[width=.4\textwidth]{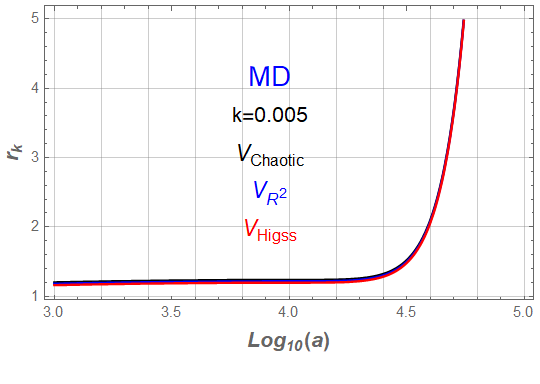}
	\caption{The numerical of $r_{k}$ in terms of $\log_{10}a$ for three different periods (inflation, RD, and MD), where we set $H_{0}=1$, $k=1$ in inflation, $k=0.01$ at RD and $k=0.005$ at MD for simplicity.}
	\label{fig: op s rk}
\end{figure}

Fig. \ref{fig: op s rk} clearly illustrates the evolution of \( r_k \) across various epochs, including inflation, RD, and MD. It is evident that the different potentials produce similar trends for \( r_k \). Unlike in Fig. \ref{fig: s rk}, there is no oscillation of \( r_k \) once dissipation effects are considered. Additionally, the trend of \( r_k \) varies, particularly during the MD phase, where the evolution consistently shows an increase. These differences will result in distinctive evolutions of Krylov complexity and Krylov entropy.

\subsection{Krylov complexity and Krylov entropy with open-system methodology}
\label{section KC in op}
In this section, we continue our investigation using the wave function \eqref{wave function of open system} to calculate the Krylov complexity and Krylov entropy. The procedure for these calculations follows precisely that of our previous study \cite{Li:2024kfm}. The only modification is the substitution of the parameter $r_k$ for $\eta$ in the formulas for Krylov complexity and Krylov entropy, which are given as follows:
\begin{equation}
K=\frac{\rm sech^{2}r_{k}|1-\mu_{1}^{2}|\tanh^{2}r_{k}}{[1+2\mu_{2}\tanh r_{k}+(\mu_{2}^{2}-|1-\mu_{1}^{2}|)\tanh^{2}r_{k}]^{2}}
\label{k complexity}
\end{equation}
and 
\begin{equation}
\begin{split}
S_{k}=&\ \frac{\rm sech^{2}r_{k}}{A^{2}}\bigl[(1+\mu_{2}\tanh r_{k})^{2}\ln(1+\mu_{2}\tanh r_{k})^{2}-\tanh^{2}r_{k}|1-\mu_{1}^{2}|\ln\tanh^{2}r_{k}\\ &\ -[(1+\mu_{2}\tanh r_{k})^{2}-|1-\mu_{1}^{2}|\tanh^{2}r_{k}]\ln \rm sech^{2} r_{k}-\tanh^{2}r_{k}|1-\mu_{1}^{2}|\ln|1-\mu_{1}^{2}| \bigr]
\end{split}
\label{k entropy}
\end{equation}
where we have defined $A = 1 + 2\mu_{2}\tanh r_{k} + (\mu_{2}^{2} - |1 - \mu_{1}^{2}|)\tanh^{2}r_{k}$. It straightforwardly demonstrates the leading order of Krylov complexity \eqref{k complexity} and Krylov entropy \eqref{k entropy} will reduce to Eq. \eqref{Krylov complexity} and Eq. \eqref{Krylov entropy} under the weak dissipative approximation, namely, $\mu_2\ll 1$.

\begin{figure}
	\centering
	\includegraphics[width=.4\textwidth]{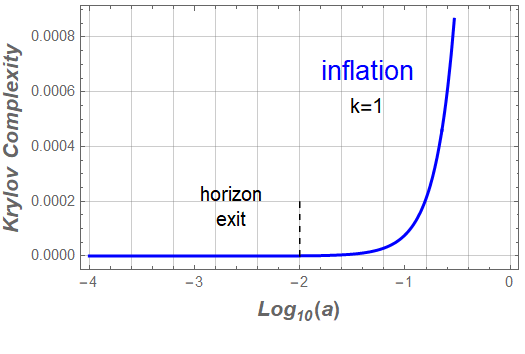}\\
	\qquad
	\includegraphics[width=.4\textwidth]{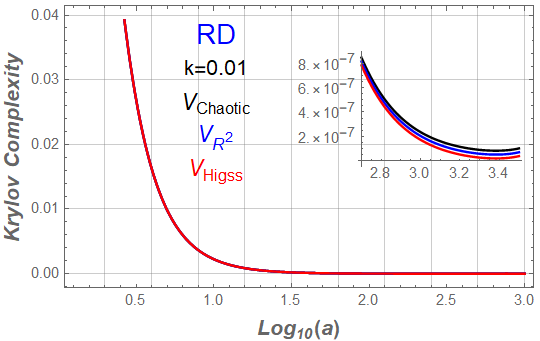}
	\qquad
	\includegraphics[width=.4\textwidth]{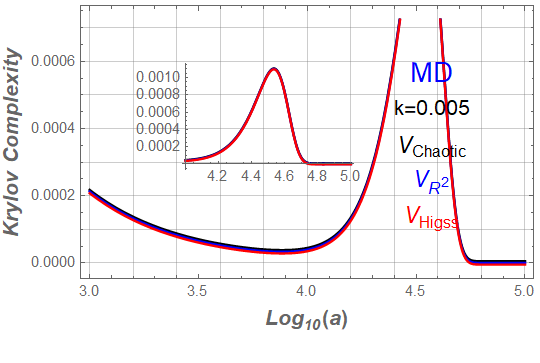}
	\caption{The numerical of Krylov complexity in terms of $\log_{10}a$ for three different periods (inflation, RD, and MD), where we set $H_{0}=1$, $k=1$ in inflation, $k=0.01$ at RD and $k=0.005$ at MD for simplicity.}
	\label{fig: op s KC}
\end{figure}

Fig. \ref{fig: op s KC} illustrates the evolution of Krylov complexity during inflation, RD, and MD. During inflation, Krylov complexity exhibits exponential growth, a behavior consistent with the closed-system methodology. In contrast, its evolution in the RD and MD epochs differs significantly with the closed-system methodology as shown in Fig. \ref{fig: s KC}. For the open-system methodology, Krylov complexity in RD shows a decreasing trend, while in MD it first peaks as the scale factor $a$ increases and subsequently declines to a constant value. Notably, the magnitude of Krylov complexity in the open-system methodology is substantially lower than in the closed-system methodology across both RD and MD.

Our numerical analysis indicates that this suppression originates from the $\rm sech(r_k)$ term in Eq. \eqref{k complexity}, an effect presents the dissipation strength. We have shown that both the RD and MD epochs can be approximated as weakly dissipative systems, characterized by $\mu_2<1$. Recalling that Krylov complexity quantifies operator growth in a many-body system, and our earlier finding $b_n=\alpha n$ indicates that the unverse is fully chaotic, the significantly reduced operator growth observed in the open-system methodology reveals that: dissipation generically leads to a faster suppression of operator growth compared to that in a closed-system methodology.

\begin{figure}
	\centering
	\includegraphics[width=.4\textwidth]{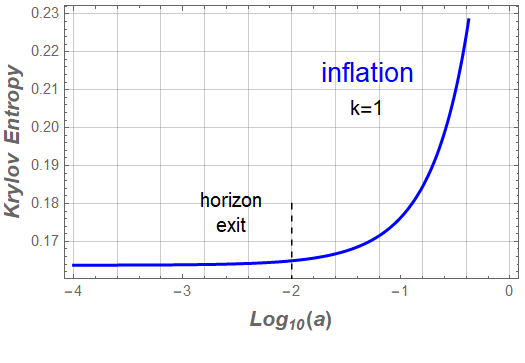}\\
	\qquad
	\includegraphics[width=.4\textwidth]{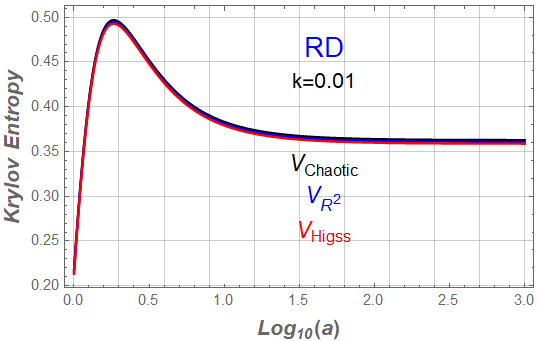}
	\qquad
	\includegraphics[width=.4\textwidth]{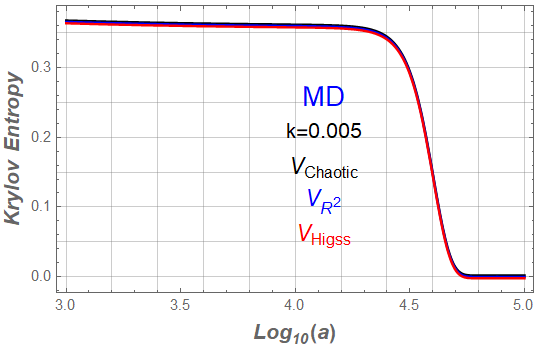}
	\caption{The numerical of Krylov entropy in terms of $\log_{10}a$ for three different periods (inflation, RD, and MD), where we set $H_{0}=1$, $k=1$ in inflation, $k=0.01$ at RD and $k=0.005$ at MD for simplicity.}
	\label{fig: op s KE}
\end{figure}

An understanding of the role of dissipation makes it clear that the magnitude of the Krylov entropy in open-system methodology is lower than that in the closed-system methodology, as demonstrated by a comparison of Figs. \ref{fig: op s KE} and \ref{fig: s KE}. The evolutionary trend is particularly distinct during the RD and MD epochs, as shown in Fig. \ref{fig: op s KE}. In RD, the Krylov entropy initially rises to a peak before saturating to a constant value. In MD, it first remains at a nearly constant level before declining and settling at a lower asymptotic value. Overall, both the cosmological background conditions and dissipative effects significantly influence the behavior of Krylov complexity and Krylov entropy. 

In this section, we employed the wave function \eqref{wave function of open system} to study the Krylov complexity and Krylov entropy. By comparing our results with those from a closed-system methodology, we observed that dissipative effects not only alter the overall evolution of the system but also suppress the magnitude of both the Krylov complexity and entropy. Since Krylov complexity quantifies operator growth, a fundamental quantum feature of dynamical systems, the suppressed growth in the open-system methodology indicates that dissipation induces more rapid decoherence-like behavior. Furthermore, we note that different inflationary potentials yield nearly identical trends for both Krylov complexity and entropy across the various cosmological epochs.

\section{Summary and outlook}
\label{section summary and outlook}

The definition of Krylov complexity is precise and unambiguous, particularly in contrast to the various definitions of circuit complexity. Its uniqueness remove the need to introduce a parametric manifold to conceptualize complexity. Furthermore, Krylov complexity characterizes the growth of operator supported within a dynamical system, and its magnitude indicates the effective quantum size of that system. Krylov complexity is fundamentally based on the Lanczos algorithm, a powerful method for diagnosing the nature of dynamical systems, whether chaotic, integrable, or free. The key diagnostic is the behavior of the Lanczos coefficient \( b_n \), whose asymptotic growth with index \( n \) serves as an indicator of the system's dynamical phase. For instance, a linear growth \( b_n \sim \alpha n \) is a hallmark of quantum chaos. In our work, we emphasize that the Lanczos coefficient explicitly reveals the early universe as a many-body, infinite, and chaotic system.
From a cosmological standpoint, the universe is fundamentally an open system. Notably, the Lanczos algorithm provides a natural framework for constructing the wave function of an open system using the second kind of Meixner polynomials. Given these insights, it is particularly important to investigate Krylov complexity throughout the evolution of the early universe, encompassing the inflationary, RD, and MD eras.

In this paper, we thoroughly examine Krylov complexity in both open- and closed-system methodologies. Our primary result is the derivation of the two-mode squeezed state using the open-system methodology, along with the calculation of its evolution equations for \(\phi_k(\eta)\) and \(r_k(\eta)\). Below, we present our key findings.

$(a)$. In Sec. \ref{Lanczos algorithm}, we introduce the Lanczos algorithm and the associated Lanczos coefficients, a framework applicable to both open and closed systems. As highlighted in \cite{Parker:2018yvk}, the Liouvillian superoperator effectively acts as a Hamiltonian when expressed in terms of creation and annihilation operators. Our investigation of Krylov complexity spans the entire evolution of the universe, including the inflationary epoch, RD and MD eras, where the slow-roll conditions are violated, the latter necessitating the inclusion of potential energy contributions.

Sec. \ref{section some basics of early universe} examines several representative inflationary potentials, such as the Higgs potential, the Starobinsky inflationary potential, and chaotic inflationary models. These models collectively cover a broad ranges of inflationary scenarios, unified through the characteristics of $\alpha$-attractors. To accurately extract the dynamical information encoded in these potentials, we analyze the evolution of the field $\phi$ relative to the scale factor during preheating, governed by Eq. \eqref{eom of phit1}. From this analysis, we compute the effective mass for the various inflationary potentials, as shown in Fig. \ref{fig:RD MD v2 v3}, where our results indicate that how the effective mass varies with respect to scale factor.

$(b)$. Considering the scale factor and various potentials, we investigate Krylov complexity using the closed-system methodology. First, we numerically simulate the parameter $r_k$
for the two-mode squeezed state, as shown in Fig. \ref{fig: s rk}. The results reveal oscillations before horizon exits and then it exponetially grows during inflation, which become enhanced in the RD and MD eras, eventually settling into constant values. Notably, different potentials yield nearly identical trends for $r_k$. Based on the numerical results for $r_k$, we also simulate the corresponding Krylov complexity and Krylov entropy, as depicted in Figs. \ref{fig: s KC} and \ref{fig: s KE}. Both Krylov complexity and entropy exhibit similar trends across the respective epochs. During inflation, we observe exponential growth, consistent with our previous study \cite{Li:2024kfm}. In the RD and MD phases, the trends display an initial enhancement followed by saturation to constant values. Moreover, various inflationary models demonstrate nearly identical trends for both Krylov complexity and entropy, confirming the approximation that the effective mass of the inflaton is approximately constant, as indicated in \cite{Li:2024ljz}.

$(c)$. Within the open-system methodology, we rigorously derive the wave function of an open system using the second kind of Meixner polynomials, dubbed as the open two-mode squeezed state. This wave function is represented in Eq. \eqref{wave function of open system}. Throughout this process, we discover that the Lanczos coefficient \(b_n\) and the dissipative coefficient $\mu_2$ are precisely determined by the specific Hamiltonian, leading to the conclusion that \(b_n\) is identical across both methodologies. Our findings indicate that \(b_n = \alpha n\), suggesting that our universe behaves as an entirely chaotic system. Consequently, the Lyapunov index is equivalent to \(b_n\) when \(n=2\). Figure \ref{fig: op s bn} illustrates the chaos present across various epochs. During the inflationary period, chaos increases exponentially before diminishing to a certain level, only to rise again due to the preheating process. Additionally, we examine the evolution of the dissipative coefficient in RD and MD. This analysis builds on previous work in Ref. \cite{Li:2024kfm}, which discusses the strong dissipative nature of the system during inflation. From Fig. \ref{fig: op s u2}, we confirm that the RD and MD systems are weakly dissipative, characterized by $\mu_2 \ll 1$. 

Additionally, we derive the evolution of the squeezing parameters $r_k$ and $\phi_k$. Their solutions are coupled, as described by Eq. \eqref{evolution in ete}, meaning the solution for $r_k$ depends on that for $\phi_k$. Using the wave function given in Eq. \eqref{wave function of open system}, we compute the Krylov complexity and Krylov entropy, presented in Eqs. \eqref{k complexity} and \eqref{k entropy}. Their leading-order forms are in good agreement with the corresponding formulas derived within the closed-system methodology.

Following the standard procedure, we require the numerical values of $r_k$ in the open-system framework, as shown in Fig. \ref{fig: op s rk}. In contrast to the closed-system case, no oscillations in $r_k$ are observed during inflation. The resulting trends for Krylov complexity and Krylov entropy are depicted in Figs. \ref{fig: op s KC} and \ref{fig: op s KE}, respectively. Compared to the closed-system results, the magnitudes of both Krylov complexity and entropy are significantly reduced. Recalling that Krylov complexity quantifies operator growth in a dynamical system, a smaller value indicates suppressed operator growth. This suppression reflects a decoherence-like behavior in the context of Krylov complexity once considering the open-system methodology.
While we have achieved a comprehensive understanding of Krylov complexity in the early universe, many ideas still remain to be explored. Here, we outline several future directions.

Our investigation begins with Krylov complexity in single-field inflation. A natural extension of this work involves generalizing it to multi-field frameworks \cite{Liu:2020zzv, Liu:2019xhn, Liu:2021rgq, Zhang:2022bde} and to theories of $f(R)$ gravity \cite{Liu:2018hno, Liu:2020zlr, Liu:2018htf}. Such extensions would allow us to explore the influence of field-space geometry and higher-order Ricci curvature terms on Krylov complexity. Our analysis is based on the Hamiltonian given in \eqref{Hamiltonian of standard case}, which is Hermitian and incorporates open-system methodology. In conventional quantum mechanics, open systems are often treated using a Lindbladian master equation \cite{Lindblad:1975ef, Gorini:1975nb}, typically involving a non-Hermitian effective Hamiltonian. In our present framework, a systematic approach involves expanding the action to higher orders in the Mukhanov-Sasaki variable.

Secondly, our findings demonstrate that in a weakly dissipative system (such as during RD or MD), the behavior of Krylov complexity deviates from the general behavior outlined in Sec.~\ref{general discussion of phi}. One plausible explanation for this discrepancy is the neglect of decoherence effects in the curvature perturbations upon horizon exit. As suggested in Ref.~\cite{Bhattacharjee:2022lzy}, decoherence can lead to a saturation phenomenon in circuit complexity. A conceptually similar mechanism may govern Krylov complexity, particularly regarding the influence of decoherence on the structure of the wave function.

Thirdly, having derived the open two-mode squeezed state wave function presented in \eqref{wave function of open system}, several promising avenues for further investigation emerge. For instance, this wave function can be used to calculate two-point correlation functions and derive the corresponding power spectrum. Given that its derivation is model-independent, it offers the flexibility to test its validity by applying it within different theoretical models. Conversely, parameters such as $\mu_1$ and $\mu_2$ can be constrained using observational data from the Cosmic Microwave Background (CMB). Furthermore, it is noteworthy that Ref.~\cite{Adhikari:2022oxr} suggests the potential applicability of the Complexity=Volume (CV) conjecture to Krylov complexity during inflation, although their analysis is confined to a closed system. In contrast, since our universe is fundamentally an open system, our framework provides a distinct opportunity to investigate the validity of the CV conjecture for Krylov complexity within an open-system methodology.

\section*{Acknowledgements}
LH and KH are funded by NSFC grant NO. 12165009, Hunan Natural Science Foundation NO. 2023JJ30487 and NO. 2022JJ40340.

\appendix
\section{The calculation of $r_k$ and $\phi_{k}$ within open-system methodology}
\label{Appendix A}
In this appendix, we present the detailed derivations and calculations for the parameters  $r_k$ and $\phi_k$ based on the open two-mode squeezed state,
\begin{equation}
|\psi\rangle=\frac{\rm{sech}\ r_{k}}{1+\mu_{2}\tanh r_{k}}\sum_{n=0}^{\infty}(-1)^{n}|1-\mu_{1}^{2}|^{\frac{n}{2}}e^{2in\phi_{k}}\Bigl (\frac{\tanh r_{k}}{1+\mu_{2}\tanh r_{k}}\Bigr )^{n}|n,n\rangle_{\vec{k},-\vec{k}}
\end{equation}
We will write down our Hamiltonian \eqref{Hamiltonian of standard case} as follows, 
\begin{equation}
\hat{H}=\bigl (k+\frac{a^{2}V_{,\phi\phi}}{2k}\bigr )(\hat{c}_{\vec{k}}\hat{c}_{\vec{k}}^{\dagger}+\hat{c}_{-\vec{k}}^{\dagger}\hat{c}_{-\vec{k}})+\bigl (\frac{a^{2}V_{,\phi\phi}}{2k}+i\frac{a'}{a}\bigr )\hat{c}_{-\vec{k}}^{\dagger}\hat{c}_{\vec{k}}^{\dagger}+\bigl (\frac{a^{2}V_{,\phi\phi}}{2k}-i\frac{a'}{a}\bigr )\hat{c}_{-\vec{k}}\hat{c}_{\vec{k}}
\end{equation}
Starting from the Schrödinger equation $\hat{H}\psi = i \partial_\eta \psi$ (with $\hbar=1$), we first evaluate its left-hand side as follows.
\begin{equation}
\begin{split}
\hat{H}|\psi\rangle = &\ \frac{\rm sech~r_{k}}{1+\mu_{2}\tanh r_{k}}\sum_{n=0}^{\infty}(-1)^{n}|1-\mu_{1}^{2}|^{\frac{n}{2}}e^{2in\phi_{k}}\Bigl (\frac{\tanh r_{k}}{1+\mu_{2}\tanh r_{k}}\Bigr )^{n}\Bigl [\bigl (k+\frac{a^{2}V_{,\phi\phi}}{2k}\bigr )(\hat{c}_{\vec{k}}\hat{c}_{\vec{k}}^{\dagger}\\ 
&\ +\hat{c}_{-\vec{k}}^{\dagger}\hat{c}_{-\vec{k}})|n,n\rangle_{\vec{k},-\vec{k}}+\bigl (\frac{a^{2}V_{,\phi\phi}}{2k}+i\frac{a'}{a}\bigr )\hat{c}_{-\vec{k}}^{\dagger}\hat{c}_{\vec{k}}^{\dagger}|n,n\rangle_{\vec{k},-\vec{k}}+\bigl (\frac{a^{2}V_{,\phi\phi}}{2k}-i\frac{a'}{a}\bigr )\hat{c}_{-\vec{k}}\hat{c}_{\vec{k}}|n,n\rangle_{\vec{k},-\vec{k}}\Bigr ]\\
= &\ \frac{\rm sech\ r_{k}}{1+\mu_{2}\tanh r_{k}}\sum_{n=0}^{\infty}(-1)^{n}|1-\mu_{1}^{2}|^{\frac{n}{2}}e^{2in\phi_{k}}\Bigl (\frac{\tanh r_{k}}{1+\mu_{2}\tanh r_{k}}\Bigr )^{n}\Bigl [\bigl (k+\frac{a^{2}V_{,\phi\phi}}{2k}\bigr )(2n+1)\\ 
&\ |n,n\rangle_{\vec{k},-\vec{k}}+\bigl (\frac{a^{2}V_{,\phi\phi}}{2k}+i\frac{a'}{a}\bigr )(n+1)|n+1,n+1\rangle_{\vec{k},-\vec{k}}+\bigl (\frac{a^{2}V_{,\phi\phi}}{2k}-i\frac{a'}{a}\bigr )n|n-1,n-1\rangle_{\vec{k},-\vec{k}}\Bigr ]\\
=&\ \frac{\rm sech\ r_{k}}{1+\mu_{2}\tanh r_{k}}\Bigl [\bigl (k+\frac{a^{2}V_{,\phi\phi}}{2k}\bigr )-|1-\mu_{1}^{2}|^{\frac{1}{2}}e^{2i\phi_{k}}\frac{\tanh r_{k}}{1+\mu_{2}\tanh r_{k}}\bigl (\frac{a^{2}V_{,\phi\phi}}{2k}-i\frac{a'}{a}\bigr )\Bigr ]|0,0\rangle_{\vec{k},-\vec{k}}\\
&\ +\frac{\rm sech\ r_{k}}{1+\mu_{2}\tanh r_{k}}\Bigl [\bigl (k+\frac{a^{2}V_{,\phi\phi}}{2k}\bigr )-|1-\mu_{1}^{2}|^{\frac{1}{2}}e^{2i\phi_{k}}\frac{\tanh r_{k}}{1+\mu_{2}\tanh r_{k}}\bigl (\frac{a^{2}V_{,\phi\phi}}{2k}-i\frac{a'}{a}\bigr )\Bigr ]\\ 
&\ \sum_{n=1}^{\infty}(-1)^{n}|1-\mu_{1}^{2}|^{\frac{n}{2}}e^{2in\phi_{k}}\Bigl (\frac{\tanh r_{k}}{1+\mu_{2}\tanh r_{k}}\Bigr )^{n}|n,n\rangle_{\vec{k},-\vec{k}}+\frac{\rm sech\ r_{k}}{1+\mu_{2}\tanh r_{k}}\Bigl [\bigl (2k+\frac{a^{2}V_{,\phi\phi}}{k}\bigr )\\
&\ -|1-\mu_{1}^{2}|^{\frac{1}{2}}e^{2i\phi_{k}}\frac{\tanh r_{k}}{1+\mu_{2}\tanh r_{k}}\bigl (\frac{a^{2}V_{,\phi\phi}}{2k}-i\frac{a'}{a}\bigr )-|1-\mu_{1}^{2}|^{-\frac{1}{2}}e^{-2i\phi_{k}}\Bigl (\frac{\tanh r_{k}}{1+\mu_{2}\tanh r_{k}}\Bigr )^{-1}\\
&\ \bigl (\frac{a^{2}V_{,\phi\phi}}{2k}+i\frac{a'}{a}\bigr )\Bigr ]\sum_{n=1}^{\infty}(-1)^{n}n|1-\mu_{1}^{2}|^{\frac{n}{2}}e^{2in\phi_{k}}\Bigl (\frac{\tanh r_{k}}{1+\mu_{2}\tanh r_{k}}\Bigr )^{n}|n,n\rangle_{\vec{k},-\vec{k}}
\end{split}
\label{eq op l-SE}
\end{equation}
Then, the right hand side is following as, 
\begin{equation}
\begin{split}
i\frac{d}{d\eta}|\psi\rangle= &\ i\frac{d}{d\eta}\Bigl [\frac{\rm sech\ r_{k}}{1+\mu_{2}\tanh r_{k}}\sum_{n=0}^{\infty}(-1)^{n}|1-\mu_{1}^{2}|^{\frac{n}{2}}e^{2in\phi_{k}}\Bigl (\frac{\tanh r_{k}}{1+\mu_{2}\tanh r_{k}}\Bigr )^{n}|n,n\rangle_{\vec{k},-\vec{k}}\Bigr ]\\
=&\ i\Bigl [-\frac{(\sinh 2r_{k}+2\mu_{2}\cosh^{2}r_{k})r_{k}'+\sinh 2r_{k}\mu_{2}'}{2\cosh^{3}r_{k}(1+\mu_{2}\tanh r_{k})^{2}}\sum_{n=0}^{\infty}(-1)^{n}|1-\mu_{1}^{2}|^{\frac{n}{2}}e^{2in\phi_{k}}\Bigl (\frac{\tanh r_{k}}{1+\mu_{2}\tanh r_{k}}\Bigr )^{n}\\
&\ |n,n\rangle_{\vec{k},-\vec{k}}\Bigr ]+i\Bigl [\frac{\rm sech\ r_{k}}{1+\mu_{2}\tanh r_{k}}\sum_{n=0}^{\infty}(-1)^{n}|1-\mu_{1}^{2}|^{\frac{n}{2}}e^{2in\phi_{k}}\Bigl (\frac{\tanh r_{k}}{1+\mu_{2}\tanh r_{k}}\Bigr )^{n}\\ 
&\ \bigl [2i\phi_{k}'+\frac{2r_{k}'}{\sin 2r_{k}(1+\mu_{2}\tanh r_{k})}+\frac{|1-\mu_{1}^{2}|'}{2|1-\mu_{1}^{2}|}-\frac{\tanh r_{k}\mu_{2}'}{1-\mu_{2}\tanh r_{k}}\bigr ]|n,n\rangle_{\vec{k},-\vec{k}}\Bigr ]\\
= &\ -i\frac{(\sinh 2r_{k}+2\mu_{2}\cosh^{2}r_{k})r_{k}'+\sinh 2r_{k}\mu_{2}'}{2\cosh^{3}r_{k}(1+\mu_{2}\tanh r_{k})^{2}}|0,0\rangle_{\vec{k},-\vec{k}}\\
&\ -i\frac{(\sinh 2r_{k}+2\mu_{2}\cosh^{2}r_{k})r_{k}'+\sinh 2r_{k}\mu_{2}'}{2\cosh^{3}r_{k}(1+\mu_{2}\tanh r_{k})^{2}}\sum_{n=1}^{\infty}(-1)^{n}|1-\mu_{1}^{2}|^{\frac{n}{2}}e^{2in\phi_{k}}\Bigl (\frac{\tanh r_{k}}{1+\mu_{2}\tanh r_{k}}\Bigr )^{n}\\
&\ |n,n\rangle_{\vec{k},-\vec{k}}+\frac{\rm sech\ r_{k}}{1+\mu_{2}\tanh r_{k}}\Bigl (-2\phi_{k}'+i\frac{2r_{k}'}{\sin 2r_{k}(1+\mu_{2}\tanh r_{k})}+i\frac{|1-\mu_{1}^{2}|'}{2|1-\mu_{1}^{2}|}\\
&\ -i\frac{\tanh r_{k}\mu_{2}'}{1-\mu_{2}\tanh r_{k}}\Bigr )\sum_{n=1}^{\infty}(-1)^{n}n|1-\mu_{1}^{2}|^{\frac{n}{2}} e^{2in\phi_{k}}\Bigl (\frac{\tanh r_{k}}{1+\mu_{2}\tanh r_{k}}\Bigr )^{n}|n,n\rangle_{\vec{k},-\vec{k}}
\end{split}
\label{eq op r-SE}
\end{equation}
Substituting Eqs. \eqref{eq op l-SE} and \eqref{eq op r-SE} into the Schrödinger equation yields the following simplified form for the open system's squeezed state:
\begin{equation}
\begin{split}
&\ A_{1}|0,0\rangle_{\vec{k},-\vec{k}}+A_{1}\sum_{n=1}^{\infty}(-1)^{n}|1-\mu_{1}^{2}|^{\frac{n}{2}}e^{2in\phi_{k}}\Bigl (\frac{\tanh r_{k}}{1+\mu_{2}\tanh r_{k}}\Bigr )^{n}|n,n\rangle_{\vec{k},-\vec{k}}\\
&\ +A_{2}\sum_{n=1}^{\infty}(-1)^{n}n|1-\mu_{1}^{2}|^{\frac{n}{2}}e^{2in\phi_{k}}\Bigl (\frac{\tanh r_{k}}{1+\mu_{2}\tanh r_{k}}\Bigr )^{n}|n,n\rangle_{\vec{k},-\vec{k}}\\
=&\ B_{1}|0,0\rangle_{\vec{k},-\vec{k}}+B_{1}\sum_{n=1}^{\infty}(-1)^{n}|1-\mu_{1}^{2}|^{\frac{n}{2}}e^{2in\phi_{k}}\Bigl (\frac{\tanh r_{k}}{1+\mu_{2}\tanh r_{k}}\Bigr )^{n}|n,n\rangle_{\vec{k},-\vec{k}}\\
&\ +B_{2}\sum_{n=1}^{\infty}(-1)^{n}n|1-\mu_{1}^{2}|^{\frac{n}{2}}e^{2in\phi_{k}}\Bigl (\frac{\tanh r_{k}}{1+\mu_{2}\tanh r_{k}}\Bigr )^{n}|n,n\rangle_{\vec{k},-\vec{k}}.
\end{split}
\end{equation}
Thus, we obtain $A_{1}=B_{1}$ as
\begin{equation}
\begin{split}
&\ -i\frac{(\sinh 2r_{k}+2\mu_{2}\cosh^{2}r_{k})r_{k}'+\sinh 2r_{k}\mu_{2}'}{2\cosh^{3}r_{k}(1+\mu_{2}\tanh r_{k})^{2}}\\ =\frac{\rm sech\ r_{k}}{1+\mu_{2}\tanh r_{k}}&\ \Bigl [\bigl (k+\frac{a^{2}V_{,\phi\phi}}{2k}\bigr )-|1-\mu_{1}^{2}|^{\frac{1}{2}}e^{2i\phi_{k}}\frac{\tanh r_{k}}{1+\mu_{2}\tanh r_{k}}\bigl (\frac{a^{2}V_{,\phi\phi}}{2k}-i\frac{a'}{a}\bigr )\Bigr ],
\end{split}
\end{equation}
and we're concerned about the imaginary component,
\begin{equation}
\begin{split}
&\ -i\frac{(\sinh 2r_{k}+2\mu_{2}\cosh^{2}r_{k})r_{k}'+\sinh 2r_{k}\mu_{2}'}{2\cosh^{3}r_{k}(1+\mu_{2}\tanh r_{k})^{2}}\\ =-\frac{\rm sech\ r_{k}}{1+\mu_{2}\tanh r_{k}}&\ \Bigl [|1-\mu_{1}^{2}|^{\frac{1}{2}}\frac{\tanh r_{k}}{1+\mu_{2}\tanh r_{k}}\bigl (\frac{a^{2}V_{,\phi\phi}}{2k}\sin 2\phi_{k}-\frac{a'}{a}\cos 2\phi_{k}\bigr )\Bigr ]. 
\end{split}
\end{equation}
and therefore, we obtain the evolution of $r_k$ as, 
\begin{equation}
r_{k}'=\frac{|1-\mu_{1}^{2}|^{\frac{1}{2}}\sinh 2r_{k}\bigl (\frac{a^{2}V_{,\phi\phi}}{2k}\sin 2\phi_{k}-\frac{a'}{a}\cos 2\phi_{k}\bigr )-\sinh 2r_{k}\mu_{2}'}{(\sinh 2r_{k}+2\mu_{2}\cosh^{2}r_{k})}.
\label{rk evolution}
\end{equation}
The other side, we have $A_{2}=B_{2}$ showing as
\begin{equation}
\begin{split}
&\ \frac{\rm sech\ r_{k}}{1+\mu_{2}\tanh r_{k}}\Bigl [-2\phi_{k}'+i\frac{2r_{k}'}{\sin 2r_{k}(1+\mu_{2}\tanh r_{k})}+i\frac{|1-\mu_{1}^{2}|'}{2|1-\mu_{1}^{2}|}-i\frac{\tanh r_{k}\mu_{2}'}{1-\mu_{2}\tanh r_{k}}\Bigr ]\\
=&\ \frac{\rm sech\ r_{k}}{1+\mu_{2}\tanh r_{k}} \Bigl [\bigl (2k+\frac{a^{2}V_{,\phi\phi}}{k}\bigr )-|1-\mu_{1}^{2}|^{\frac{1}{2}}e^{2i\phi_{k}}\frac{\tanh r_{k}}{1+\mu_{2}\tanh r_{k}}\bigl (\frac{a^{2}V_{,\phi\phi}}{2k}-i\frac{a'}{a}\bigr )\\
&\ -|1-\mu_{1}^{2}|^{-\frac{1}{2}} e^{-2i\phi_{k}}\Bigl (\frac{\tanh r_{k}}{1+\mu_{2}\tanh r_{k}}\Bigr )^{-1} \bigl (\frac{a^{2}V_{,\phi\phi}}{2k}+i\frac{a'}{a}\bigr )\Bigr ], 
\end{split}
\end{equation}
the real component reads as,
\begin{equation}
\begin{split}
-\frac{2\phi_{k}'\rm sech\ r_{k}}{1+\mu_{2}\tanh r_{k}} &\ =\frac{\rm sech\ r_{k}}{1+\mu_{2}\tanh r_{k}}\Bigl [\bigl (2k+\frac{a^{2}V_{,\phi\phi}}{k}\bigr )\\
-\bigl (\frac{a^{2}V_{,\phi\phi}}{2k}\cos 2\phi_{k}+\frac{a'}{a}\sin 2\phi_{k}\bigr )&\ \Bigl (|1-\mu_{1}^{2}|^{\frac{1}{2}}\frac{\tanh r_{k}}{1+\mu_{2}\tanh r_{k}}+|1-\mu_{1}^{2}|^{-\frac{1}{2}}(\tanh^{-1}r_{k}+\mu_{2})\Bigr )\Bigr ]. 
\end{split}
\end{equation}
Finally, we obtain the evolution equation of $\phi_k$ as follows, 
\begin{equation}
\begin{split}
\phi_{k}' =&\ -\bigl (k+\frac{a^{2}V_{,\phi\phi}}{2k}\bigr )+\frac{1}{2}\bigl (\frac{a^{2}V_{,\phi\phi}}{2k}\cos 2\phi_{k}+\frac{a'}{a}\sin 2\phi_{k}\bigr )\\
&\ \Bigl [|1-\mu_{1}^{2}|^{\frac{1}{2}}\frac{\tanh r_{k}}{1+\mu_{2}\tanh r_{k}}+|1-\mu_{1}^{2}|^{-\frac{1}{2}}(\tanh^{-1}r_{k}+\mu_{2})\Bigr ]. 
\label{phik evolution}
\end{split}
\end{equation}
It can be verified that Eqs. \eqref{rk evolution} and \eqref{phik evolution} correctly reduce to Eq. \eqref{evolution in standard case} in the limit $\mu_1 = \mu_2 = 0$.

\end{document}